\documentstyle[epsf]{mnjer}
\title[Luminosity Function Evolution by Spectral Type]
{Autofib Redshift Survey: II -- The Evolution of the Galaxy Luminosity
Function by Spectral Type}

\author[J.Heyl et al.]{
{\LARGE \rm Jeremy Heyl$^{1,2}$\thanks{Please direct correspondence
to: Jeremy Heyl, Lick Observatory, University of California, Santa Cruz, CA~95064, USA}, Matthew Colless$^3$, Richard S.\
Ellis$^1$ and Tom Broadhurst$^{4,5}$} \\
$^1$Institute of Astronomy, Madingley Road, Cambridge CB3 0HA, UK \\
$^2$Lick Observatory, Board of Studies in Astronomy and Astrophysics,
    University of California, Santa Cruz, CA~95064, USA \\
$^3$Mount Stromlo and Siding Spring Observatories, The Australian
    National University, Weston Creek, ACT 2611, Australia \\
$^4$Department of Physics and Astronomy, Johns
    Hopkins University, Baltimore MD~21218, USA \\
$^5$Department of Astronomy, University of California, Berkeley,
    CA~94720, USA } 
\date {Accepted ---. Received ---; in original form ---.}

\def\plotone#1{\centering \leavevmode
\epsfxsize=\columnwidth \epsfbox{#1}}
\def\plottwo#1#2{\centering \leavevmode
\epsfxsize=.45\columnwidth \epsfbox{#1} \hfil
\epsfxsize=.45\columnwidth \epsfbox{#2}}
\def\plottwow#1#2{\centering \leavevmode
\epsfxsize=.45\textwidth \epsfbox{#1} \hfil
\epsfxsize=.45\textwidth \epsfbox{#2}}
\def\plotfiddle#1#2#3#4#5#6#7{\centering \leavevmode \vbox
to#2{\rule{0pt}{#2}}
    \includegraphics{\ifepsfdraft}}
\def\plotsmlandmn#1{\plotfiddle{#1}{0.75\columnwidth}{270}{32}{32}{-130}{175}}

\def\rmmat#1{{\hbox{\rm #1}}}
\def\rmscr#1{\rmmat{\scriptsize #1}}
\def\vm{$1/V_\rmscr{max}$}
\def\vvm{$V/V_\rmscr{max}$}
\newcommand{\be}{\begin{equation}}
\newcommand{\ee}{\end{equation}}
\newcommand{\comment}[1]{}
\newcommand{\bt}{\begin{table} \begin{center}}
\newcommand{\et}{\end{center} \end{table}}

\def\gs{\mathrel{\raise0.3ex\hbox{$\scriptstyle >$}\kern-0.70em
\lower0.71ex\hbox{{$\scriptstyle \sim$}}}}
\def\ls{\mathrel{\raise0.3ex\hbox{$\scriptstyle <$}\kern-0.70em
\lower0.71ex\hbox{{$\scriptstyle \sim$}}}}
\def\magn{{\rm\,mag}}

\def\Otwo{[O{\sc II}]}
\def\unsetyr{\def\oyear{\relax}\def\cyear{\relax}}
\def\setyr{\def\oyear{(}\def\cyear{)}}
\unsetyr
\def\poy{\protect\oyear}
\def\pcy{\protect\cyear}
\def\jcite#1{\setyr\cite{#1}\unsetyr}

\begin{document}
\textheight=8.5truein

\label{firstpage}
\maketitle

\begin{abstract}
We determine the evolution of the galaxy luminosity function (LF) as a
function of spectral type using the Autofib redshift survey, a
compendium of over 1700 galaxy redshifts in various magnitude-limited
samples spanning $b_J$=11.5--24.0. To carry out this analysis we have
developed a cross-correlation technique which classifies faint galaxy
spectra into one of six types based on local galaxy templates. Tests
and simulations show that this technique yields classifications
correct to within one type for more than 90\% of the galaxies in our
sample. We have also developed extensions of the step-wise maximum
likelihood method and the STY parametric method for estimating LFs
which are applicable to recovering an evolving LF from multiple
samples. We compare these methods to the standard \vm\ method used in
Paper~I.

Applying these spectral classifications and LF estimators to the
Autofib sample, we find that: (i)~The spectra and LF of E/S0 galaxies
show no appreciable evolution out to at least $z$$\sim$0.5.
(ii)~Early-type spirals show modest evolution, characterised by a
gradual steepening of the faint end of their LF with redshift.
(iii)~Out to $z$$\sim$0.5, the overall evolution of the galaxy
population is dominated by changes seen in late-type
spirals. The characteristic luminosity ($L^*$) of these galaxies 
appears to brighten with redshift and there are signs of strong
density evolution (a rapid increase in $\phi^*$). These effects
appear to be luminosity dependent so that the LF steepens at
higher redshift. These trends are accompanied by a steep increase
in the median \Otwo\ equivalent width, implying a rapid increase in
the star-formation rate with redshift at fixed luminosity---a given
star-formation rate is found at higher redshift in galaxies of higher
luminosity. We find that these conclusions are robust with respect to
spectral classification errors and the luminosity function estimator.
Finally, we briefly discuss the correlations between our ground-based
dataset and a subset of 60 galaxies for which WFPC-2 images are
now available from Hubble Space Telescope.

\end{abstract}


\section{Introduction}

The Autofib Redshift Survey was introduced in the first paper in this
series (\cite{Elli96}, Paper~I). It is an extensive compilation of
over 1700 galaxy redshifts drawn from 53 separate magnitude-limited
slices spanning 11.5$<$$b_J$$<$24.0. About half of the data has been
published by our observing team during 1985--1994 in various papers
(\cite{Pete85}, \cite{BES}, \cite{LDSS1}, 1993, \cite{Glaz95a}), and a
further 1026 new redshifts have been secured in the intermediate
magnitude range 17$<$$b_J$$<$22 using the Autofib fibre positioner on
the 3.9m Anglo-Australian Telescope.  The resulting dataset has the
advantage of covering a wide luminosity range up to redshifts
$z$$\approx$0.75, and is thus ideally suited to studying the evolution
of the galaxy luminosity function (LF). Details of the photometric
selection, observing techniques, spectroscopic analysis and a
comprehensive catalogue will be given in a subsequent paper
(Paper~III). We assume $H_0$=100$h$\,km\,s$^{-1}$\,Mpc$^{-1}$ and
$q_0$=0.5 throughout.

The main goal of this paper is the determination of the evolution of
the luminosity functions (LFs) by galaxy type. A number of studies at
low redshift have investigated the LFs for different morphological
types, for different colour classes, or for different degrees of
star-formation activity (indicated by \Otwo\ equivalent width). Among
the most extensive of these studies are those of \jcite{Bing88},
\jcite{Love92}, \jcite{Marz94} and \jcite{Lin96}.  This work has been
directed toward understanding differences in the shape of the LF
between the morphological types, and differences in the LFs of field
and cluster galaxies.

In the last few years the size of deep redshift surveys has increased
to the point where it has begun to be feasible to derive the evolution
of the LF with redshift. Initial analyses were carried out by
\jcite{Eale93}, \jcite{Lons93} and \jcite{Trey94} using the surveys of
\jcite{Pete85}, \jcite{BES}, \jcite{LDSS1} and \jcite{Cowi91}. These
studies clearly indicated evolution of the LF with redshift (as had
already been apparent from comparison of the galaxy number counts and
redshift distributions with non-evolving models). However the size of
these surveys was not large enough for the authors of the LF analyses
to reach a clear concensus on the form of this evolution. Two new deep
redshift surveys, the Canada-France Redshift Survey of 730 I-selected
galaxies (\cite{Lill95}) and our Autofib survey of over 1700
B-selected galaxies (\cite{Elli96}; Paper~I), have increased the
sample sizes to the point that it is now possible not only to recover
the evolution of the overall LF with redshift, but also to examine the
evolution of the LF as a function of galaxy type.

\jcite{Lill95} measured redshifts for a sample of galaxies with I$<$22
that reaches out to $z$$\sim$1, obtaining LFs as function of both
redshift and colour. They find that the LF evolution is strongly
differential with colour, in the sense that the LF of red galaxies
evolves little if at all out to $z$$\sim$1 while the blue population
shows substantial evolution.  Lilly et~al.'s work probes to the highest
redshifts yet explored, however their results are limited by the fact
that they are working with a single magnitude-limited sample, so that
there is little overlap in the luminosity range spanned in different
redshift ranges. Thus their results show different parts of the LF at
different redshifts---the faint end at low redshifts, the knee at the
median redshift of the survey, and the bright-end cutoff at
$z$$\sim$1---making it difficult to form a coherent picture of changes
in the overall shape of the LF.

The Autofib survey addresses this problem, inherent in
magnitude-limited samples, albeit at the expense of a lower mean redshift
than that of CFRS, by combining several surveys with different
magnitude limits, so that we cover a broad range of luminosities at
all redshifts out to $z$$\simeq$0.75 (see Figure~7 of Paper~I). Another
difference between this work and that of Lilly et~al.\ is that whereas
their sample is selected in the I band, and so emphasises galaxies
dominated by older stellar populations, ours is B-selected, and so
emphasises the galaxies with ongoing star-formation.  Finally, whereas
Lilly et~al.\ use colour as the indicator of galaxy type (dividing
their sample into red and blue classes), and whereas in Paper~I we
used \Otwo\ equivalent width (comparing the LFs for galaxies with EWs
greater and less than 20\AA), in this paper we use well-defined
spectral types determined directly from our survey observations.

The results from the Autofib survey, whilst confirming the basic
evolutionary trends identified by Lilly et al, are more suggestive of
the possibility of two distinct populations. In particular, the
evolving shape of the LF is more clearly seen, although to somewhat
lower redshift than for CFRS, and the trend appears to be largely due
to an increase in the volume density of 0.1 - 1 $L^{\ast}$ galaxies
with strong [O II] emission.  In Paper I, the LF was primarily explored
as a function of redshift for the entire galaxy population.  Here, we
will utilise the spectral classifications to describe the evolution of
galaxy luminosity function in greater detail. This is valuable because
the spectra themselves offer a glimpse into details of the evolving
components that cannot be gleaned from photometric and redshift data
alone. As well as using the spectra to isolate different subsets of
the data, we will coadd the spectra according to various selection criteria
in order to see if the short-term star formation histories of distant 
galaxies, as revealed by diagnostic absorption lines, are different from 
those of their local contemporaries. This is appropriate because
the Autofib spectra were obtained at unusually good spectral resolution 
for a faint sample (4-8 \AA\ for $b_J<$22 and 12 \AA\ for 22$<b_J<$24 
c.f. $\simeq$25\AA\ for the CFRS survey). For a limited subset of our 
fainter galaxies, we also briefly discuss Hubble Space Telescope (HST) 
images and correlations  between morphology and photometric classifications.

Our secondary goal in this paper is to discuss these spectral
classification techniques and also the extensions to LF estimation
methods we have developed in order to deal with the very general
nature of the survey sample and the extra degree of freedom
represented by LFs which evolve with redshift. In Paper~I these
techniques were only briefly mentioned in discussing new results on
both the faint end of the local LF and the evolution of the overall
galaxy LF since redshifts $z$$\sim$0.75.

A plan of the paper follows. The spectral classification technique and
the tests and simulations carried out to verify its precision and
reliability are described in \S\ref{sec:specclass}. Here we also
discuss the correlation between photometric properties and HST morphologies.
In \S\ref{sec:lfest} we review existing methods of LF estimation and
derive extensions to two methods which allow both a
clustering-insensitive direct estimate and a best-fit parametric model
to be recovered for an evolving LF. These methods are tested and
compared to the standard \vm\ LF estimator in \S\ref{sec:lfesttest}.
In \S\ref{sec:typelf} we use the spectral classification and LF
estimation techniques to determine the LFs as a function of both
redshift and spectral type. These results, and their implications for
the physical processes driving galaxy evolution, are discussed in
\S\ref{sec:discuss}. Our conclusions are presented in
\S\ref{sec:conclude}.

\section{Spectral Classification}
\label{sec:specclass}

In Paper~I we stressed the importance of reliable $k$-corrections for
the determination of the luminosity function over a wide redshift
range.  The lack of morphological classifications or colours for most
of the surveys combined here means that traditional methods for
estimating $k$-corrections cannot be uniformly applied.  Only the DARS
survey has been classified morphologically and only the LDSS-1 and
LDSS-2 samples have $b_j-r_f$ colours.  Applying a mean $k$-correction
or defining a new passband probably would not be fruitful as the
galaxies span $z$$\approx$0--1, so an extremely blue or red galaxy at
high redshift would have its luminosity incorrectly estimated by over
a magnitude, and using a mean-redshift could lead to errors nearly as
large. Furthermore, the volume weighting implied would be even more
uncertain. As a major goal of our survey is to estimate the LF as a
function of spectral class, we need an approach that is both uniformly
applicable and more reliable.

In principle, a $k$-correction could be `read' directly from each
flux-calibrated spectrum. In practice, the calibrations are too
uncertain to allow the continuum shape of the spectra to be matched to
a standard SED, as is demonstrated by a comparison of synthetic
colours estimated from the spectra with those obtained from direct
imaging. However, on smaller scales where spectral features can be
measured, the data are considerably more reliable. Accordingly, we
have developed a technique for classifying each galaxy spectrum in the
survey which succeeds very well for but the most noisy spectra in the
sample (those from the LDSS-1 and LDSS-2 surveys, for which we used
the $b_j-r_f$ colour to estimate the $k$-correction as in
\jcite{LDSS1}). This technique was briefly outlined in Paper~I
(\S3.1); here it is described in detail and validated by a variety of
tests.

\subsection{The Cross-Correlation Method}

To determine the spectral classification of each of the galaxies, we
chose to cross-correlate the survey spectra against those of the
\nocite{Kenn92a,Kenn92b} Kennicutt (1992a, 1992b) spectral
library. This library is highly appropriate for this task, not only in
spectral resolution and wavelength range but also because the spectra
represent the light integrated over a large portion of the local
galaxies rather than of the central regions as is usually the
case. The physical apertures involved are closely matched to those of
Autofib and LDSS at their mean redshifts.  The 2-arcsec effective
aperture of these surveys corresponds to a physical size of 2-8
$h^{-1}$ kpc for redshifts between 0.1 and 1.1, well matched to the
physical apertures used by Kennicutt (1992a, 1992b).

The cross-correlation technique works as follows. First both the 
Kennicutt template and survey spectra are smoothed on a scale of 
100\AA\ in the observer's frame and these smoothed versions are 
subtracted yielding continuum-subtracted spectra. The spectra are
then rebinned to 2\AA\ per bin in the rest frame of the galaxy.
The cross-correlation is defined to be
\be
r = { \left ( \sum_i t_i o_i \right )^2 \over \sum_i t_i^2 \sum_i o_i^2 }.
\ee
where $t_i$ and $o_i$ are flux values in the 2\AA\ bins in the
template and observed spectra respectively.  This procedure is similar
to the method derived independently by \jcite{Zari95}.  The survey spectrum is
classed according to the morphological type of the template with which
it most strongly correlates.  Each Kennicutt template has an
associated morphological classification given in Kennicutt (1992b).
We find the \jcite{King85} $k$-correction that most closely
corresponds to the Kennicutt classification, thus determining the
$k$-correction for each galaxy over a range of redshifts for which it
might be detected within the particular sub-survey given the apparent
magnitude limits. We cannot use the Kennicutt data to directly
calculate the $b_J$ $k$-corrections for $z>0.05$, as they do not
extend into the ultraviolet below 3650\AA. On the other hand, the
spectral energy distributions (SEDs) of \jcite{Penc76} and
\jcite{King85}, which {\em do} extend into the UV, lack the
small-scale spectral features that make the cross-correlation method
possible.  \jcite{Kinn93} have published a UV spectroscopic atlas
of star-forming galaxies; unfortunately, the uncertainties in these
data are too large in the wavelength range critical to estimating
$k$-corrections (2800--3700$\AA$) for this atlas to be useful for our
purposes.  Additionally, the IUE aperature is not well-matched to the
distant population.
By using the Kennicutt spectra as templates and the
appropriate SEDs for the matched types, the $k$-correction can be
readily found.

If one assumes that the noise level of the survey spectra is uniform
with wavelength (a good assumption for the faintest spectra), the
cross-correlation coefficients may be directly converted 
to a value of $\chi^2$:
\be
     \chi^2 = {1 \over \sigma^2} \sum_i o_i^2 ( 1 - r )
\ee
Clearly the best template match will yield a minimum $\chi^2$ value. 
In principle the $\chi^2$ estimator enables the method to be made more 
sophisticated by including both a distribution of prior probabilities 
among the template galaxies and a list of those templates which are a
near-equal match. This information may provide the basis of an error 
estimate for the $k$-correction of each galaxy, but is beyond the
scope of this analysis, for which we used the `raw' $r$-values to
determine the best-matching galaxy.

\subsection{Simulated Tests}
\label{sec:kc}

To verify the algorithm, we performed two series of tests. The first
involved simulating the procedure using fluxed spectra. We randomly
selected a Kennicutt spectrum, normalised it to a particular mean
number of counts per bin and added a sky signal. Using the total count
per bin, a random Gaussian deviate was chosen and this noise added
prior to sky-subtraction.  By repeating this process 100 times, we
created an ensemble of test spectra of known signal/noise ratio in
each pixel. Each of these spectra was processed as above and the
success rate in returning the correct Kennicutt class was calculated.
As Table~\ref{tab:kcsuccess} shows, the success rate is encouraging.

\bt
\caption{Success rates of the classification algorithm.}
\label{tab:kcsuccess}
\begin{tabular}{ccccc}
& \multicolumn{2}{c}{Fluxed} & \multicolumn{2}{c}{Unfluxed} \\
$\overline{S/N}$ & Template & Type    & Template & Type    \\ 
                 & Correct  & Correct & Correct  & Correct \\
4.05 & 100\% & 100\% & ~81\% & ~84\% \\
2.03 & ~94\% & ~94\% & ~76\% & ~83\% \\ 
1.22 & ~88\% & ~90\% & ~62\% & ~72\% \\
0.82 & ~82\% & ~86\% & ~52\% & ~68\% \\ 
\end{tabular}
\et

Next we tried to simulate the unfluxed spectra in the catalogue and 
also took into account the effect of varying the redshifts of the
simulated spectra. Test spectra were generated as before except 
that each spectrum was multiplied by a response function which is zero
outside 3600--7200\AA\ and increases quadratically by a factor of
two from the edge to the central wavelength. This response function can
be blueshifted by a factor of $\times$1-1.6 equivalent to a redshift
of 0--0.6. Table~\ref{tab:kcsuccess} shows the success rates for
these tests.

\begin{figure}
\plotone{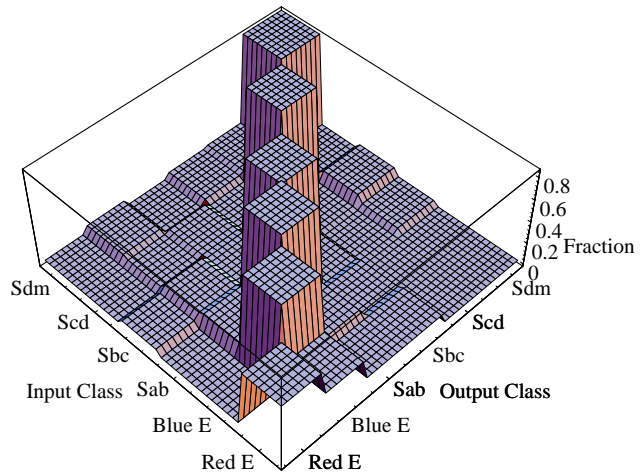}
\caption{Input versus output classifications.  The height of the
surface represents the fraction of objects within each input class
that were classified into each output. Red ellipticals and early
spirals can be confused with blue ellipticals; and intermediate
spirals can be confused both with earlier and later types.}
\label{fig:inout}
\end{figure}

Figure~\ref{fig:inout} examines the results of the ``unfluxed''
simulations in more detail. The most striking feature of the
distribution is the diagonal ridge line which traces the correct
classifications.  Classes~\#2~(red ellipticals) and~\#4 (early
spirals) are sometimes confused with class~\#3 (blue ellipticals).
Furthermore, class~\#5 (intermediate spirals) are sometimes classified
as class~\#3 (blue ellipticals) and as class~\#7 (star-bursting
spirals). However, the overall agreement is excellent.

As mentioned above, each $k$-correction class corresponds to several 
Kennicutt spectra, each with varying strengths of spectral features.
Consequently, with the addition of noise, the spectra can be confused 
across a $k$-correction class. In the case of spectra for intermediate 
spirals which contain weak features of both earlier and later classes, 
the classification may be significantly in error.

\begin{figure}
\plotone{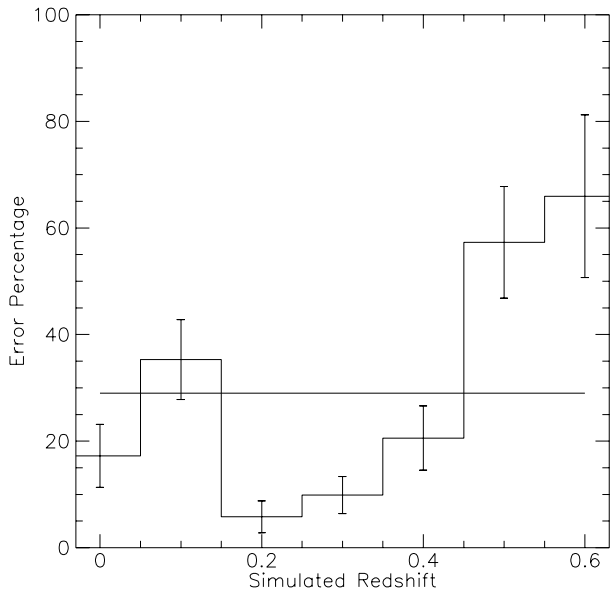}
\caption{Classification error rate as a function of redshift. The
error increases markedly with redshift, as the identifying features of
the spectra are lost off the red end of our simulated spectrograph.
The horizontal line is the mean error rate of the various tests: 29\%.}
\label{fig:kcerror}
\end{figure}

Finally, we performed the blueshifting described above to simulate how 
observing galaxies at various redshifts through a fixed wavelength range can
affect the success rate. Figure~\ref{fig:kcerror} shows that the error 
rate increases because a decreasing portion of a galaxy's spectrum overlaps
with the Kennicutt templates. We can only expect a satisfactory classification
to $z\simeq0.6$. For higher redshift galaxies in the
LDSS-1 survey, and for all the LDSS-2 survey galaxies, we use colour
to determine spectral class.

\subsection{Sample Spectra}

\begin{figure}
\def\jplotsmlandmn#1{\plotfiddle{#1}{0.7\columnwidth}{270}{32}{32}{-130}{175}}
\jplotsmlandmn{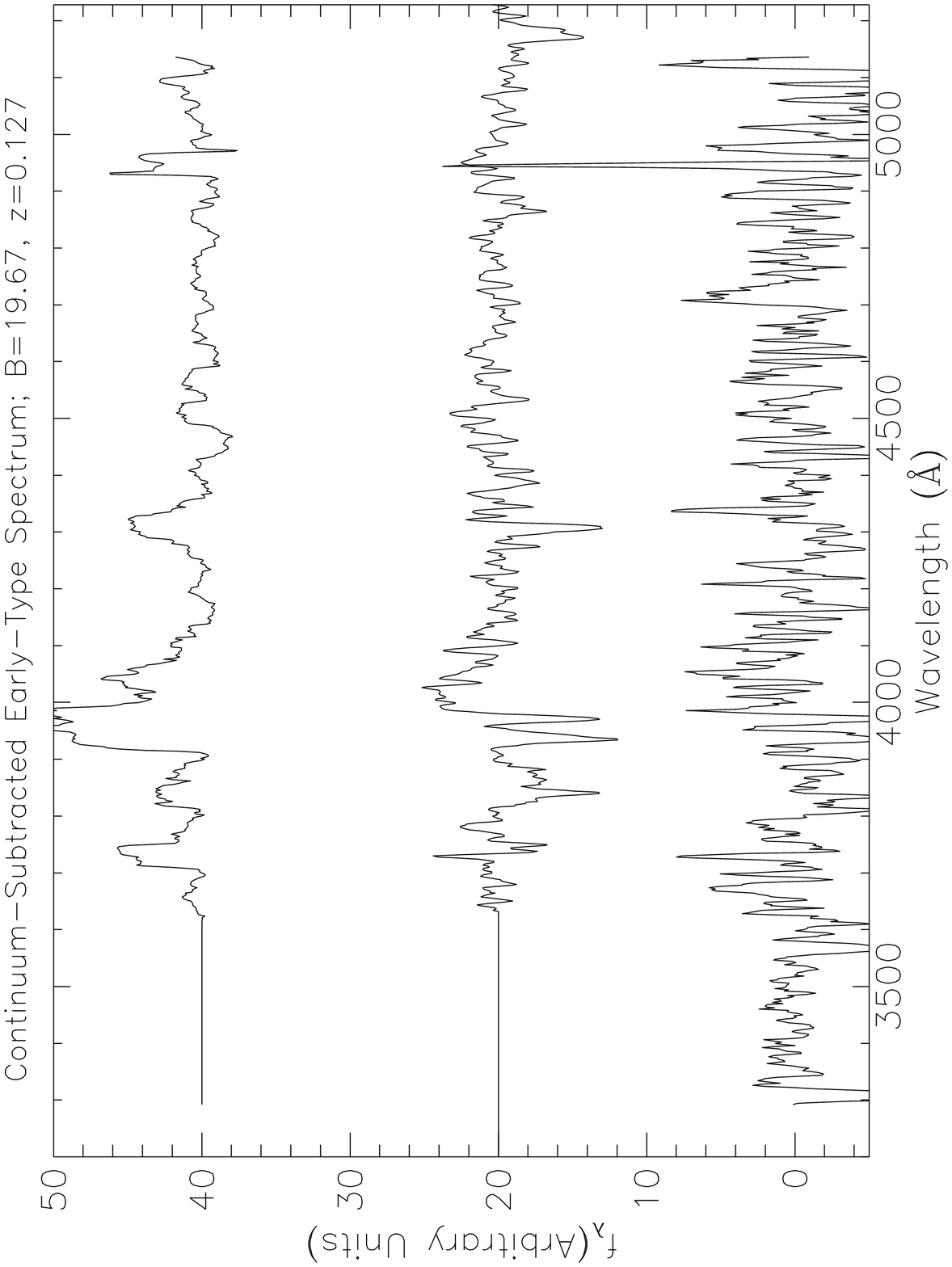} \\
\jplotsmlandmn{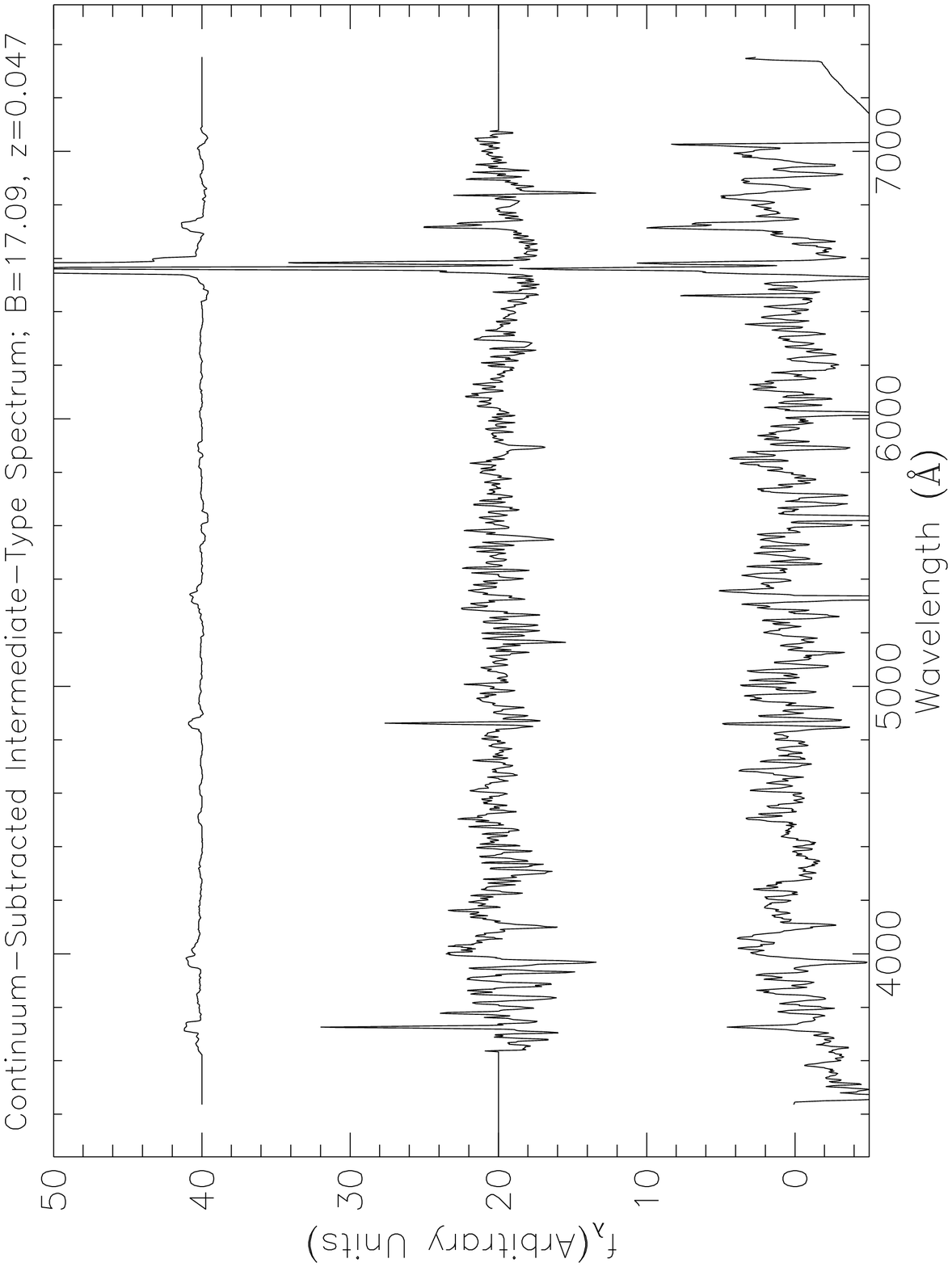} \\
\jplotsmlandmn{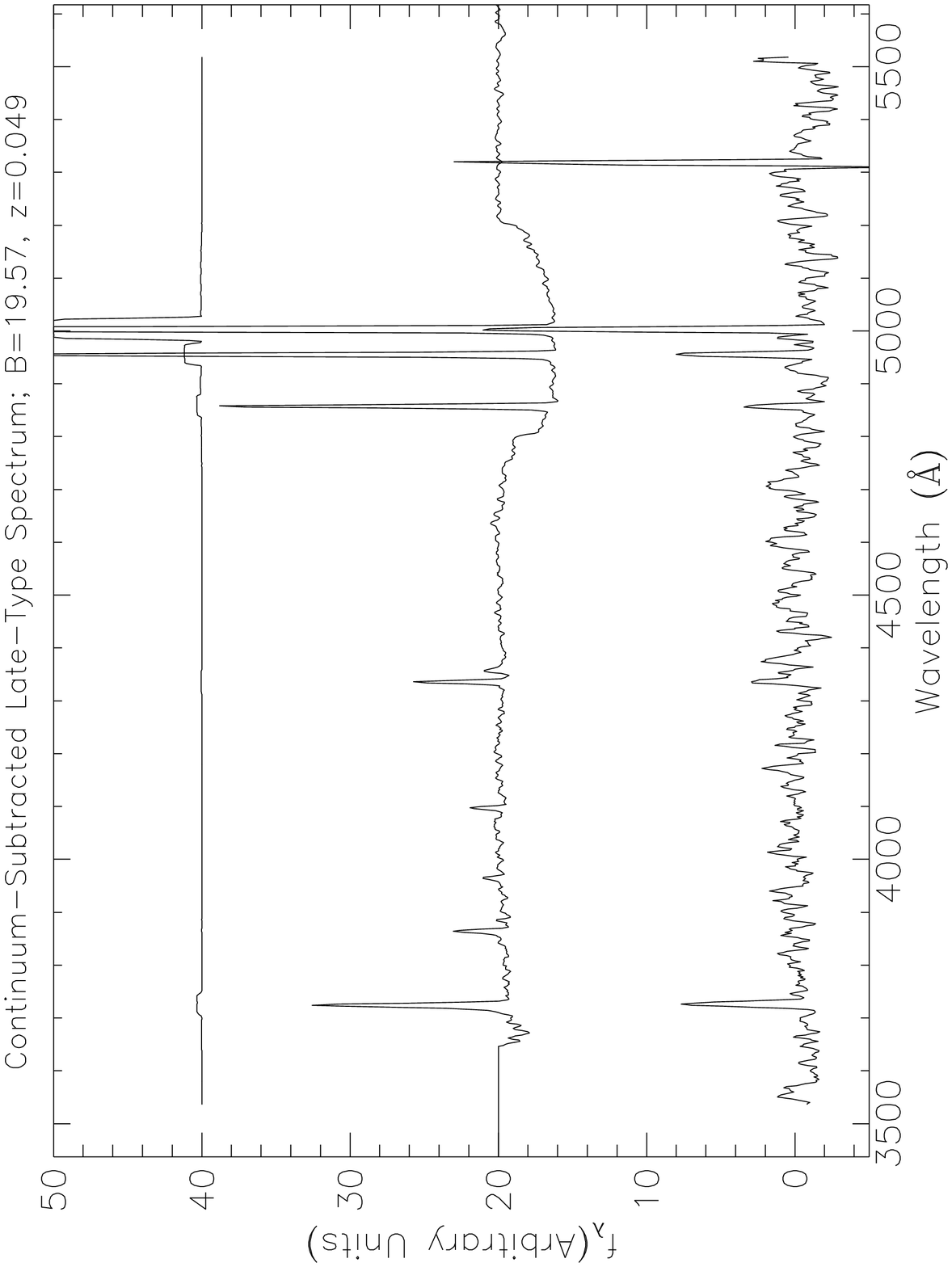}
\caption{Classifying bright spectra. Three continuum-subtracted spectra
selected from the bright section of the Autofib survey---from top to
bottom, an early-type galaxy (B=19.67, $z$=0.127), an intermediate-type
galaxy (B=17.09, $z$=0.047) and a late-type galaxy (B=19.57, $z$=0.049).
 The lower curve in each panel traces the the observed spectrum after
the continuum has been subtracted. best-fitting (continuum-subtracted)
spectrum from the Kennicutt atlas. The middle curve is best-fitting
(continuum-subtracted) spectrum from the Kennicutt atlas. and the upper
is the cross-correlation contribution as a function of wavelength.}
\label{fig:egbright}
\end{figure}

\begin{figure}
\plotsmlandmn{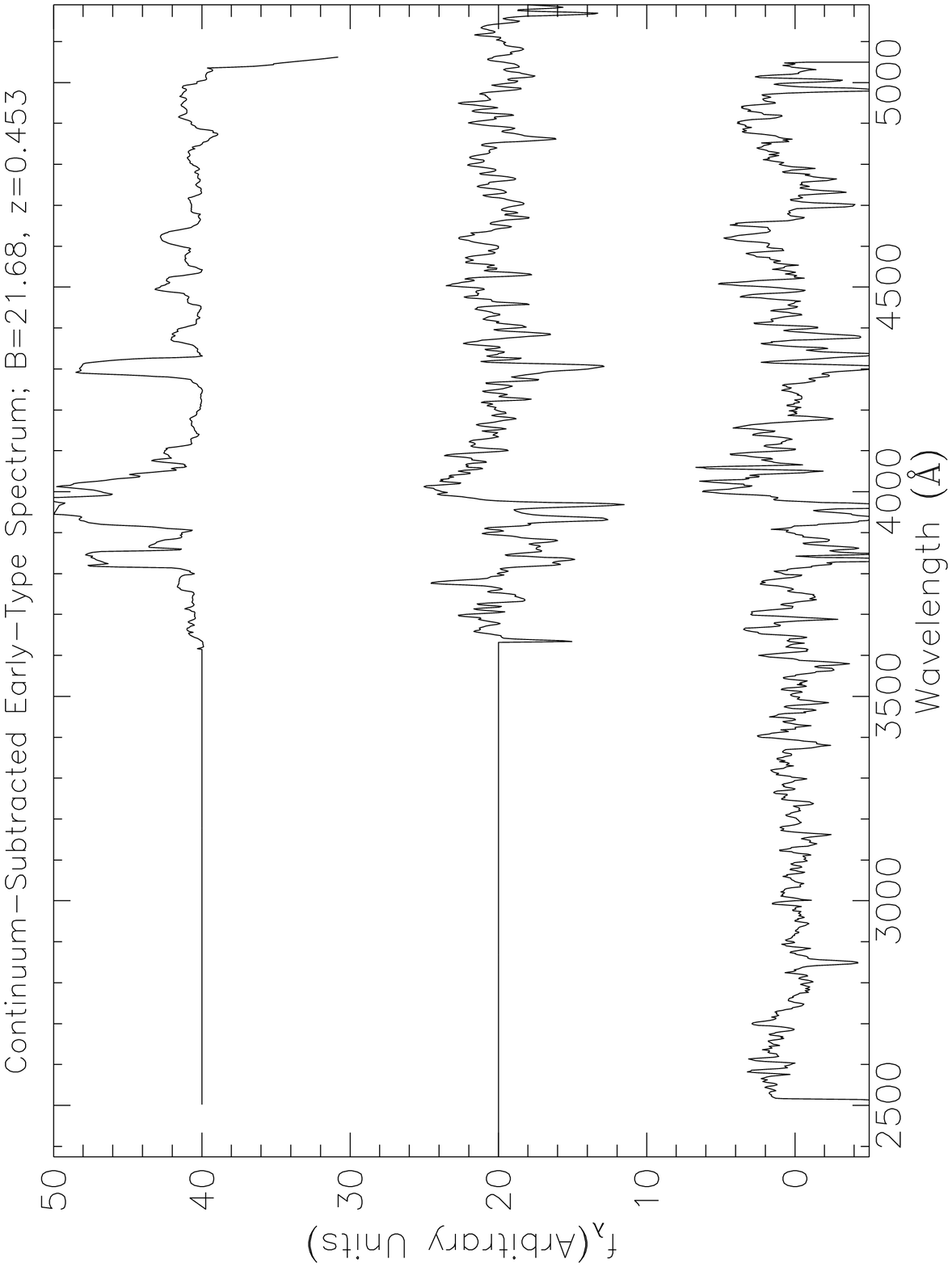}\\
\plotsmlandmn{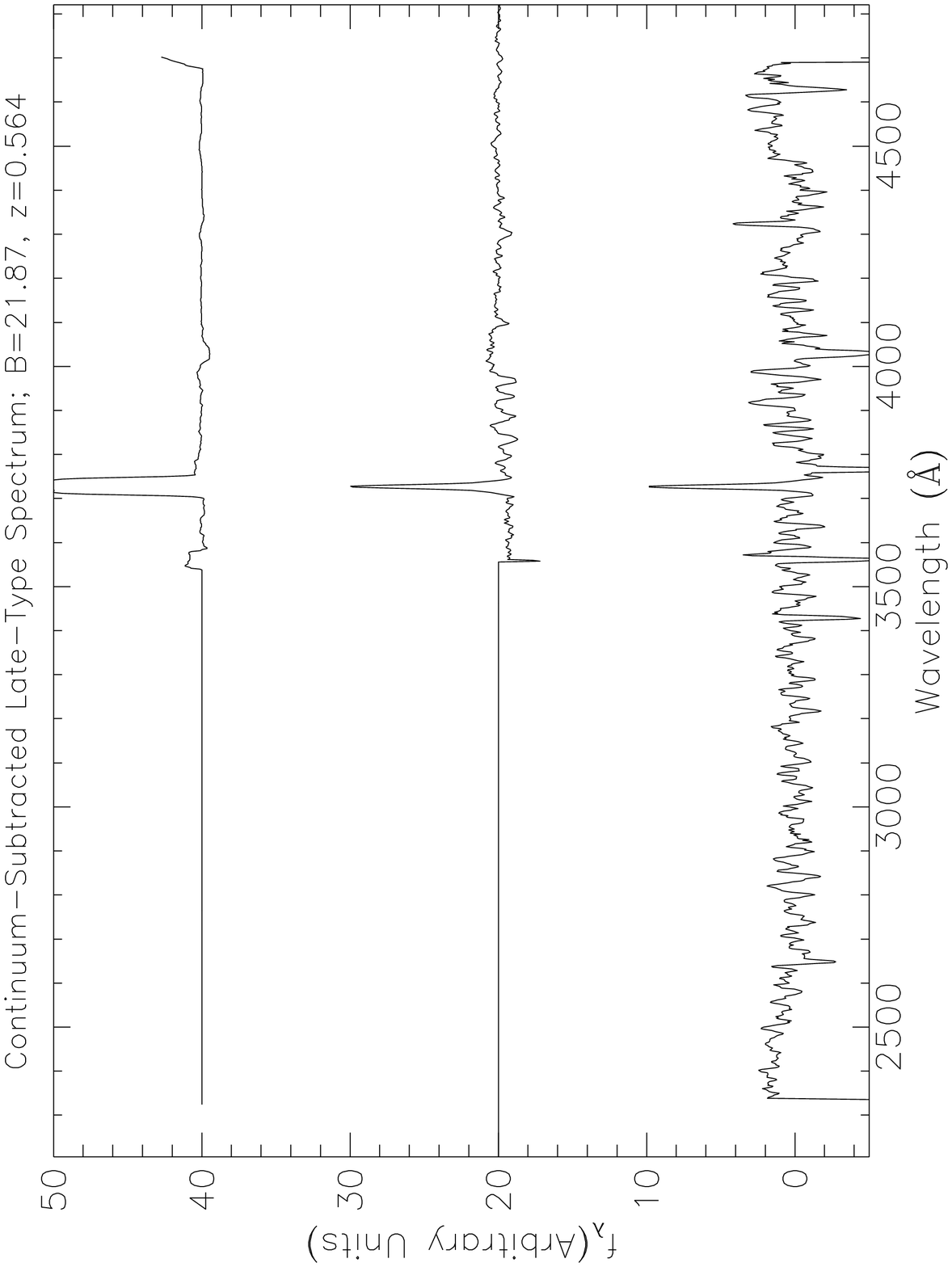}
\caption{Classifying faint spectra. Two spectra selected from the
faint section of the Autofib survey.  The upper panel is an early-type
galaxy (B=21.87, $z$=0.564). The lower panel is a late-type galaxy
(B=21.68, $z$=0.453). The arrangement is as in the previous figure.}
\label{fig:egfaint}
\end{figure}

Figures~\ref{fig:egbright} and \ref{fig:egfaint} illustrate examples
of the spectral classification technique. In each case the lower curve 
is the observed spectrum and the middle curve is the Kennicutt spectrum 
selected as the closest match. The continuum has been subtracted from 
both spectra as described earlier and the observed spectrum has been 
smoothed further to accentuate the features. The upper curve is the 
product of the two spectra smoothed over 20 bins to show which 
features contribute most strongly to the total cross-correlation. 

Examining the brighter survey galaxies (Figure~\ref{fig:egbright})
there is strong correspondence between the observed spectra and the
best templates. Intermediate-type galaxies are the most challenging as
seen from the middle panel of Figure~\ref{fig:egbright}. However even
with faint galaxies (Figure~\ref{fig:egfaint}) the algorithm performs
well. However, the limited overlap at higher redshifts is a stumbling
block (as predicted by Figure~\ref{fig:kcerror}). Presently there is no
equivalent spectroscopic atlas covering near-ultraviolet wavelengths,
and in any case there are few strong features blueward of the [OII]
emission line that would be helpful.  For more distant sources, the
most promising option would be to extend the observing window redwards
utilising night sky suppression techniques to limit the deleterious
effects of the OH emission.

\subsection{Tests with Real Spectra}
\label{sec:realtest}

A subset of the galaxies in the catalogue have $b_J-r_F$ colours and
this provide an independent test of the classification algorithm. The 
LDSS-1 galaxies (\cite{LDSS1}) lie within 21$<B<$22.5 and are amongst the
faintest in the survey. 

\begin{figure}
\plotone{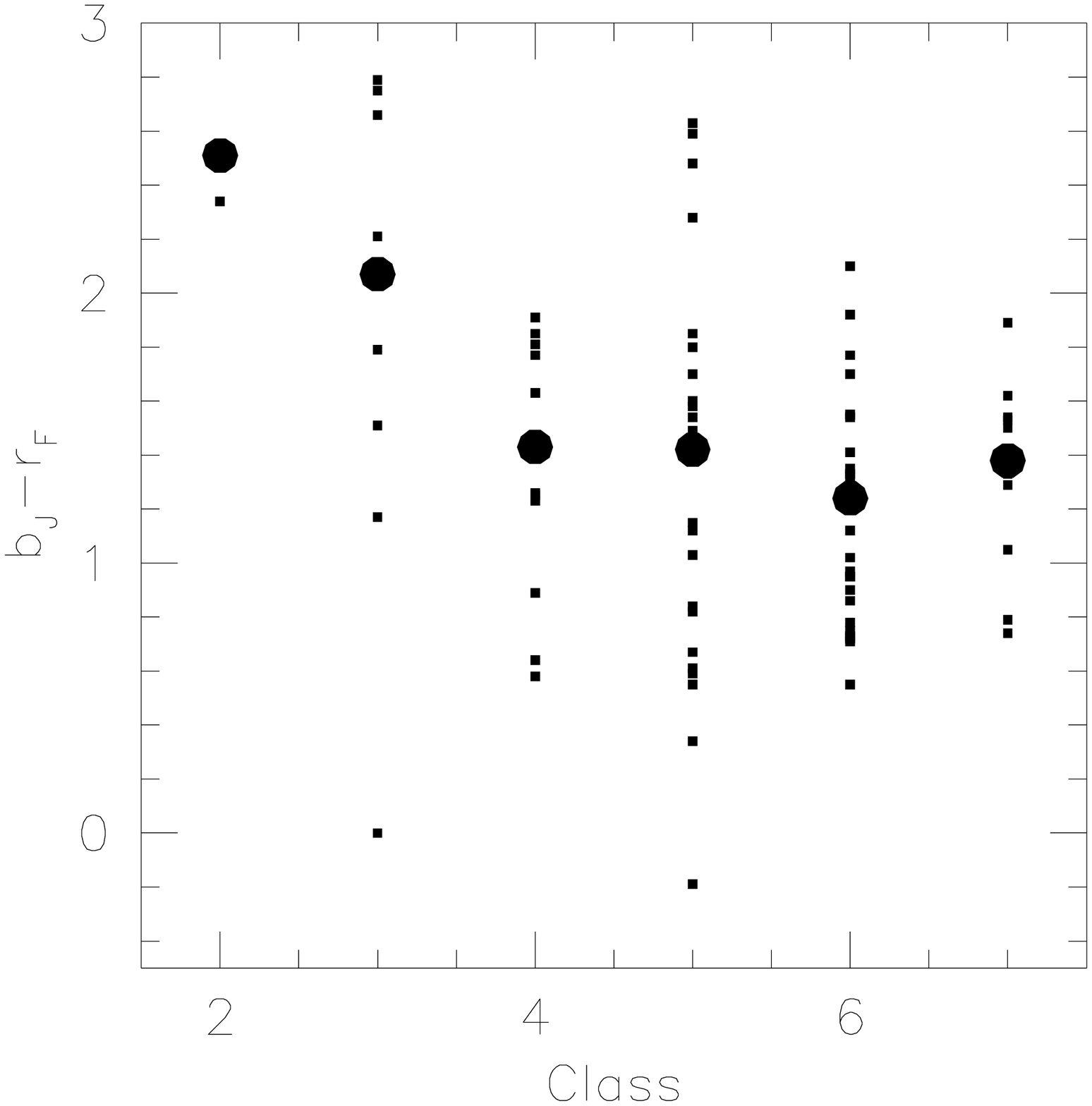}

\plotone{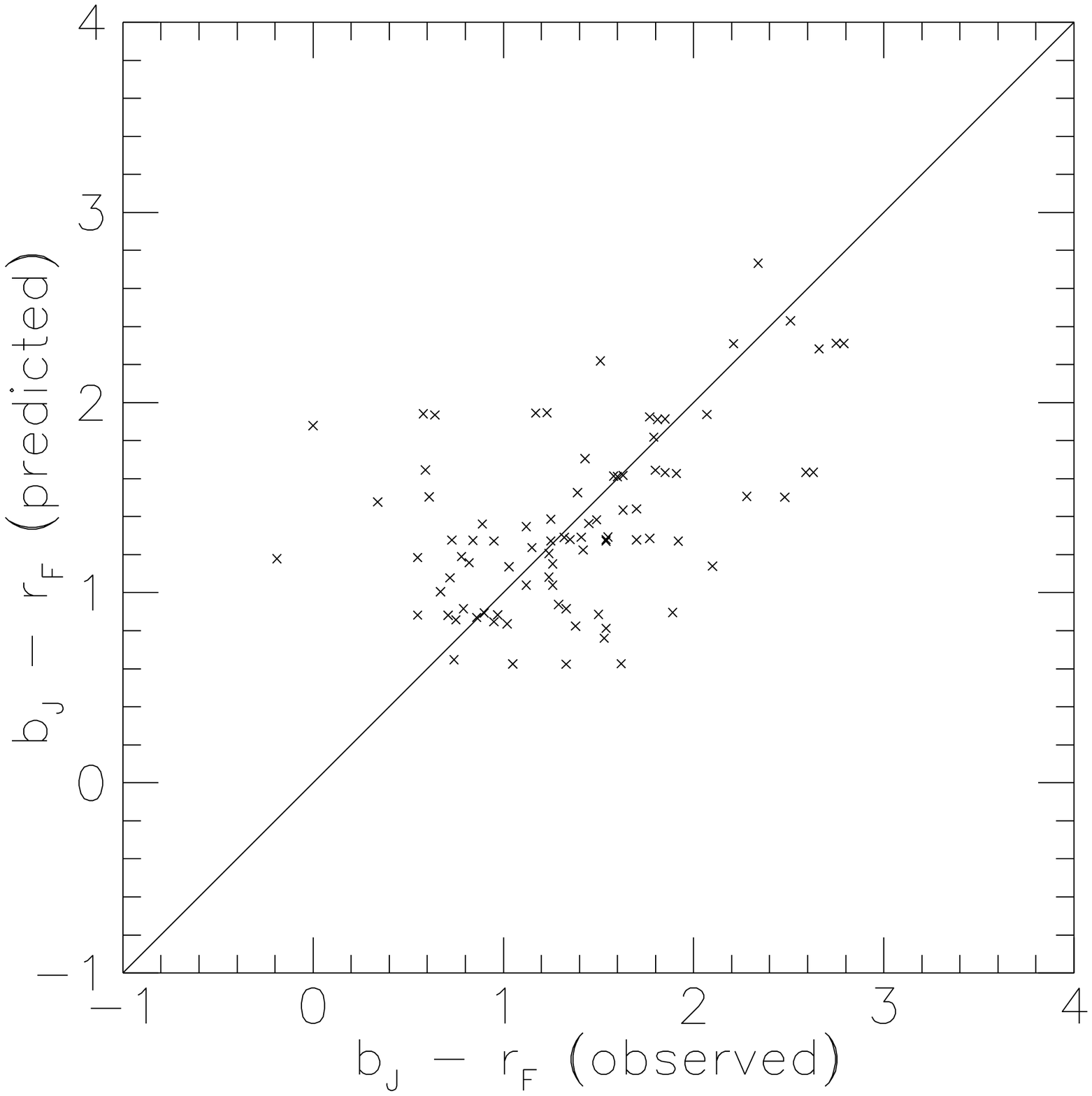}
\caption{Results for LDSS-1 spectra. The upper panel depicts the spread
in observed $b_J-r_F$ colour for the galaxies in the LDSS-1 survey
against the $k$-correction classification.  Each small square
represents a single galaxy. The large circles show the trend in the
median colour for each class. The lower panel shows colour predicted
from the $k$-correction classification at the redshift of each galaxy
against the observed colour of the galaxy. Each square represents a
galaxy in the survey.}
\label{fig:ldsscol} 
\end{figure}

We use a two-stage test: the correlation of observed colour against
$k$-correction class and correlation of `predicted' colour against
observed colour. We calculate the $b_J-r_F$ colour in the observer's 
frame by using the classification of each LDSS spectrum using our
method and then determining the colour of the appropriate SED at the 
redshift of the galaxy. Figure~\ref{fig:ldsscol} shows the results of 
both these tests. 

The trend of colour versus class (first panel) exhibits a large spread 
which could arise from observing errors, colour corrections and  
misclassifications. The second panel shows the predicted colour against
the observed colour thereby removing the effect of redshift. 
Although the trend is remarkably good the mean-absolute error is
approximately $0.4\magn$. Budgeting the errors for this multistage
process is difficult, as they arise in several places.  The $b_J$ and
$r_F$ magnitudes each have on average an uncertainty of $0.1\magn$
leading to a total error in the colour of $0.15\magn$. Furthermore, a
few galaxies have colours bluer than any of the $k$-correction SEDs; this
is worrisome, but the effect is small and limited to only a few
galaxies.  

The remaining spread of approximately $0.3\magn$ may be due to either
the variance of the intrinsic colours of galaxies within a given
spectral class (\cite{deVa77}) or to misclassifications. Although it
depends on $k$-correction class, the error in the $k$-correction is
approximately 1.5 times greater than the colour error, so if all the
spread in this plot were due to misclassifications, the $k$-corrections for
these galaxies would have an RMS error of $0.6\magn$.  The LDSS-1 data
are the most noisy of the survey and we expect some scatter in the
intrinsic galaxy colours; consequently, we expect the mean
misclassification rate and the $k$-correction error to be much lower
over the entire survey and an unimportant source of error in the
luminosity function estimates.

\subsection{Testing with HST}
\label{sec:hsttest}

Given the limited usefulness of comparing spectral classifications with those
based on broad band colours, it is interesting to compare both classifications
with those based on morphologies derived from the Hubble Space Telescope
(HST) images. For the brighter redshift surveys (\cite{Pete85}, \cite{Love92})
$k$-corrections have normally been assigned on the basis of a Hubble 
type determined from visual inspection and thus it is important, where 
possible, to examine the correlation between Hubble type and the 
diagnostics we have employed in this work to classify our spectra and
faint galaxies. 

As part of a major programme to morphologically resolve the faint field 
population, we have imaged about 80 galaxies in the fields sampled
by our survey with $21<b_J<$24 with the HST WFPC-2. Typical integration 
times were 4-6 orbits in the F814W filter. Of these 80, 64 have spectra 
and redshifts as part of the Autofib survey. Most of these are drawn 
from the LDSS-2 catalogue although a few LDSS-1 and fibre sample galaxies 
were also observed. At these apparent magnitudes the spectra are generally 
too faint or highly redshifted for the cross-correlation technique to be a reliable classifier (c.f. Figure 2) and the bulk of the classifications
are thus based on $b_J$-$r_F$ colours.
 
Each of the 64 galaxies has been assigned a morphological type by one of us
[RSE] according to the morphological scheme introduced for analyses of 
the Medium Deep Survey (\cite{Glaz95b}) and Hubble Deep Field (\cite{Abra96}).
Briefly, galaxies were classified visually on the F814W image according to
a scheme whose internal resolution adapts to the varying quality
and size of the individual images. Although the scheme has 12 one-dimensional
classes, in this comparison we have collated the data into 5 broader classes,
namely stellar/compact, spheroidal, early/intermediate spiral, late-type
spiral/Irregular, merger/unclassed. In the following we examine how well 
these HST morphologies correlate with the classes used in the 
rest of the paper noting that these are, largely, based on colour.

\begin{figure}
\plotone{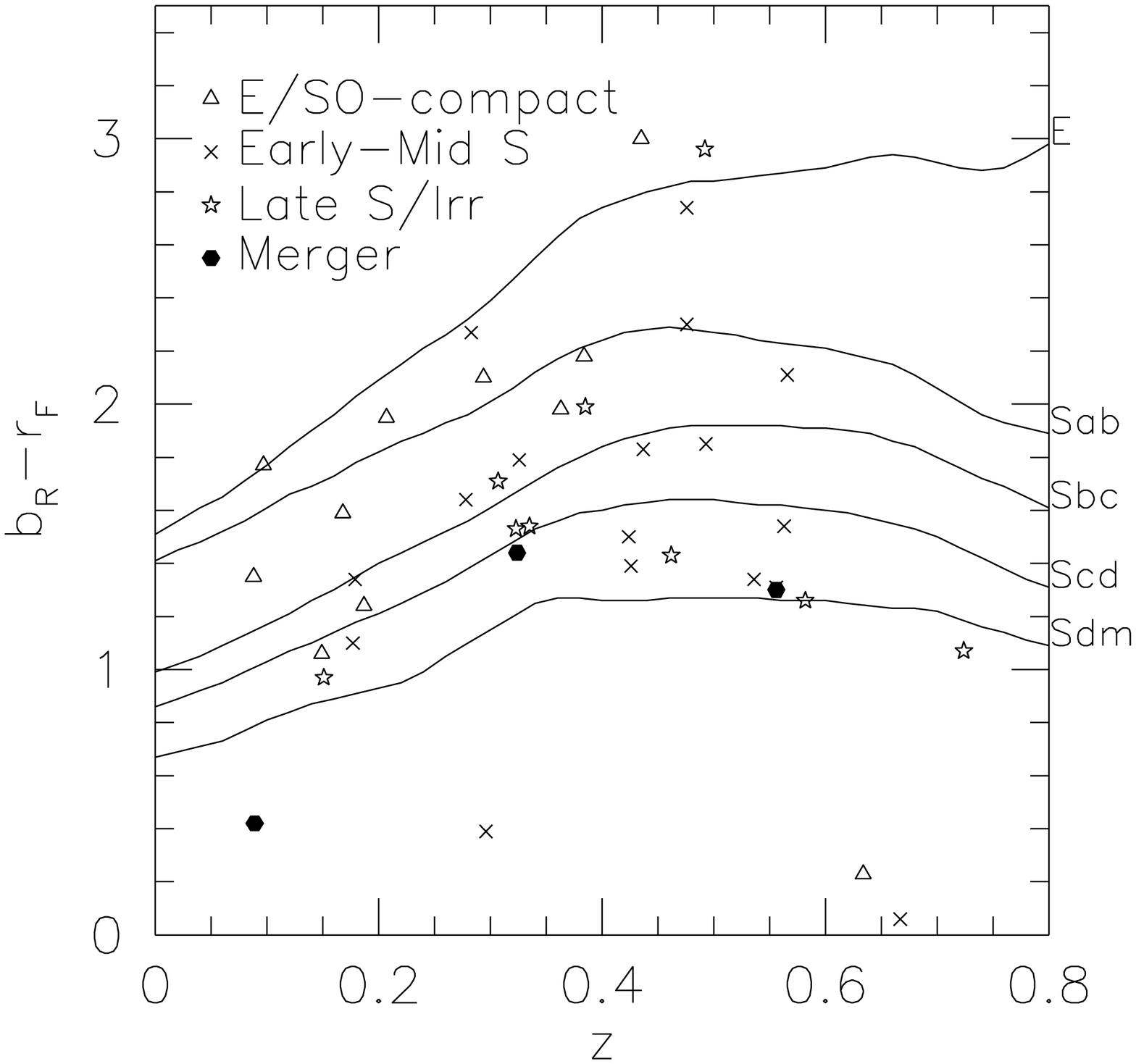} \\
\plotone{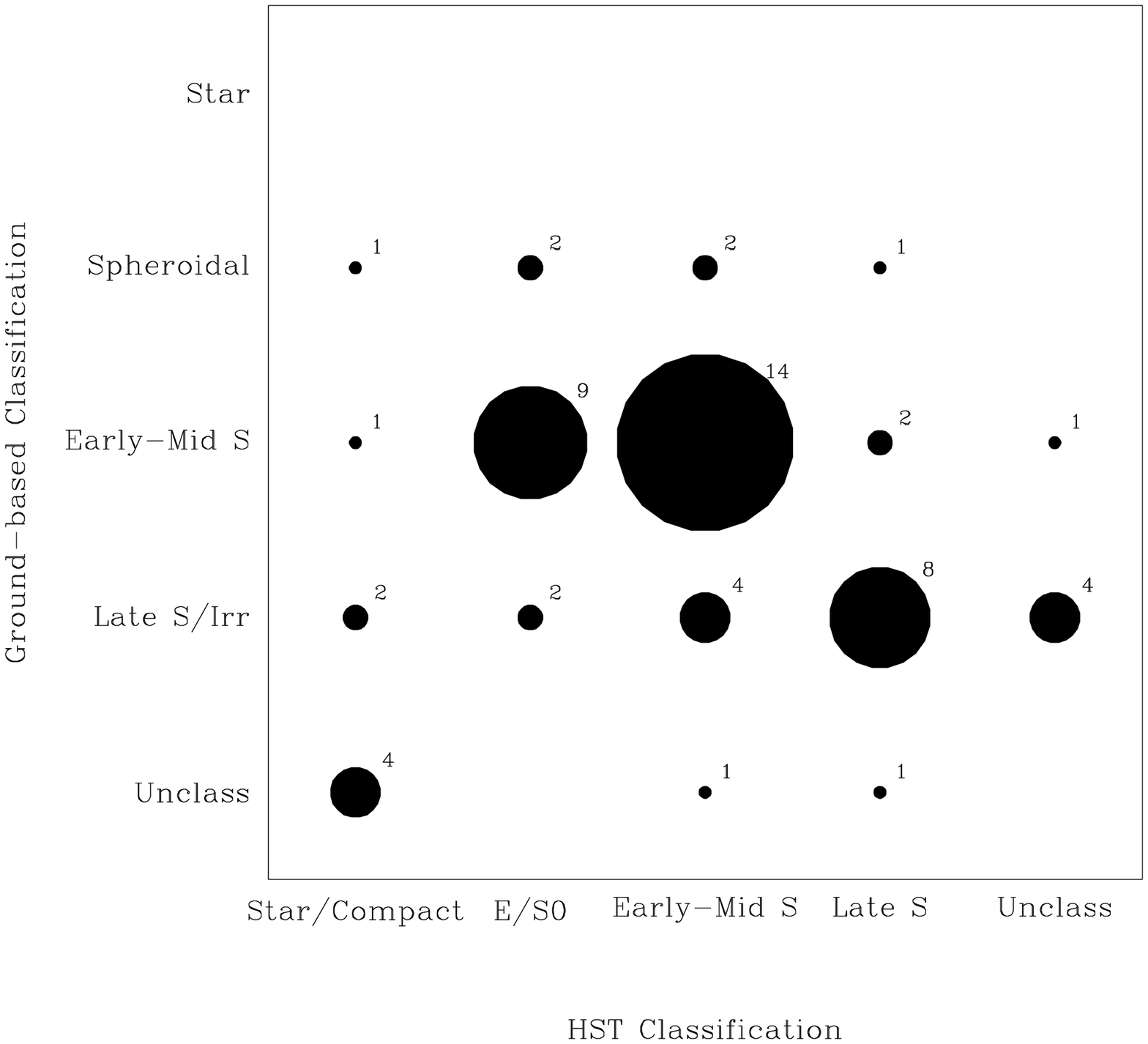}
\caption{Correlations with HST morphology. The upper panel shows the 
observed $b_J-r_F$ colour-redshift relation for 55 galaxies in the 
LDSS-1 and LDSS-2 surveys. Different symbols have been used according to
the 5 HST morphologies discussed in the text (see inset). The curves 
represent the expected trends for the adopted set of $k$-corrections. 
The lower panel shows a direct comparison between HST morphology 
and photometric/spectroscopic class including those sources which
defied classification in one or both systems. The area of each circle 
is proportional to the sample size in each category and is indicated
numerically.}
\label{fig:hstclass} 
\end{figure}

Figure~\ref{fig:hstclass} shows the $b_J$-$r_F$ colour-redshift relation
with each datapoint denoted according to its HST morphology, as well as
a direct comparison of the HST morphological type with the $k$-correction 
class based on the broad-band colours or spectroscopic cross-correlation. 
In these plots 3 QSOs in the HST sample have been removed but
objects which defied classification in one or both schemes have been
included.

As before, recognising that morphology and colour/spectral class do not
perfectly align even with high signal/noise local data, we can consider the
agreement to be satisfactory; 52\% of the HST-classed galaxies lie on
the ridge line, sharing identical classifications. The largest
systematic deviation appears to be with the morphologically-defined
spheroidal galaxies where a significant proportion have spectral or
photometric properties which class them as early-type spirals. On closer
inspection these sources include galaxies which appear to lie at the
boundary of the S0 and early-type spiral region. However, as discussed
later, there do appear to be a number of cases where HST-classed ellipticals
have slightly bluer colours than expected and this is accompanied by 
[O II] emission, so the blurring of classes in the spheroidal/early-type 
spiral region is partially a physical effect.

More extreme misclassifications than those which transfer data to
adjacent bins are rarer and, upon inspection, easier to understand. For 
example, there are quite a few compact objects which are HII-galaxies 
with extremely blue colours and spectra indicative of late-type spirals.
Under our simple morphological scheme, these are binned with the ellipticals
since this classification boundary for distant objects is difficult to 
discern. There is also one compact red object with an emission line.
 
\section{Luminosity Function Estimators}
\label{sec:lfest}

Once the luminosities of survey galaxies have been determined (see
\S~3.2 of Paper I), the next logical step in our analysis is to
construct the luminosity function as a function of class.  A number of
possible LF estimators have been discussed in the literature (see
reviews by \jcite{Felt76}, \jcite{EEP}). In the particular case of the
Autofib survey, we are interested in an estimator which is well-suited
to the deep pencil-beam nature of the survey which may imply greater
large scale structure fluctuations than would be the case for panoramic
all-sky surveys.

In Paper I, we presented LF results based on a simple estimator, the
\vm\ method (\cite{Schm68}), arguing that although the redshift
distribution in individual pencil beams may suggest large departures
from that expected for a homogeneous distribution, the large number of
individual beams (53 in 9 widely distributed regions) should minimise
such difficulties. Here we will explore this assumption more carefully
by developing a modified version of the Step-Wise Maximum-Likelihood
(SWML) method. That method, and its derivatives, has become a popular
tool to overcome the effects of fluctuations by assuming the LF has
the same {\it shape} across all density fluctuations in the
survey. Although useful in local surveys, as we are interested in
examining potential evolutionary changes in the LF this restriction
is unacceptable.  In the following sections we derive and compare
various alternate methods for estimating the LF which do not impose
this constraint.

\subsection{The 1/Vmax Method}

We begin by briefly reviewing the \vm\ method used in Paper I.
The first step in this method is the calculation of the total volume 
within which the object could have been observed. In a single 
magnitude-limited survey (see \S~3.2 of Paper I) :
\be
V_\rmscr{max} = {c \over H_0} A \int_{z_\rmscr{min}}^{z_\rmscr{max}}
{ d_L^2 dz \over (1+z)^3 \sqrt{1+2 q_0 z}} 
\ee
where $z_\rmscr{min}$ and $z_\rmscr{max}$ are the minimum and
maximum redshifts from which the object could have been observed in
the survey considering the distance modulus and $k$-correction. $A$ is
the area of sky surveyed in steradians. In its simplest form, the
luminosity function is obtained by collecting the sources in bins of
constant magnitude and summing the $(V_\rmscr{max})^{-1}$ values in
each group.

\jcite{Avni80} describe how to combine more than one sample coherently
in a \vvm\ analysis.  The new variable, denoted $V_a$ is simply the
sum of $V_\rmscr{max}$ over all the surveys in which the object could
have been observed.  Finally, \jcite{Eale93} describes how to use this
variable to construct the luminosity function as a function of redshift. 
The analysis proceeds according to Schmidt's method with the
exception that the data is also binned in redshift. Here $z_\rmscr{min}$ 
is the minimum redshift at which the object could have been observed in the
magnitude-limited sample and be in the redshift range of interest. The
method works similarly for the maximum redshift.

As mentioned, these variants of the traditional \vm\ estimator are
each affected by the presence of clustering in the sample. Clustering
will lead to a poor estimate of the faint-end slope of the luminosity
function.

\subsection{Clustering-Insensitive Methods}

Several authors have introduced estimators, which although slightly
biased, reduce the effects of clustering on the resulting LFs. Each
of these techniques attempts to maximise the likelihood of observing 
a particular set of galaxies by varying parameters of the luminosity
and completeness functions. The methods differ in the way the 
assumed probability of observing a particular galaxy is calculated.

Adopting Poisson statistics is an obvious way of proceeding. This 
assumption was first applied to the analysis of quasar samples 
(\cite{Mars83}) and later recast to analyse the CfA redshift 
survey (\cite{Chol86}). The probability of observing $k$ galaxies 
in the interval $dM dz$ in an area of sky $d\Omega$ is 
\be
P_k = e^{-\lambda} {\lambda^k \over k!}~.
\ee
where
\be
\lambda = {1 \over n} \phi(M) \rho(z) dM dz d\Omega \ee
and $n$ is the average number density of the survey.  By binning
$\phi(M)$ and $\rho(z)$, these two functions may be estimated without
assuming particular forms. The only task that remains is to maximise
the total probability of the survey (the product of all the $P_k$ in each
of the bins) while varying the values of $\phi(M)$ and $\rho(z)$.

The C-method (\cite{Lynd71}) adopts a different approach. By considering
the plane of redshift {\it versus} absolute magnitude, this method uses
the fact that the ratio of the number of galaxies observed between $L$
and $L+dL$ to the number brighter than $L$ is proportional to the
ratio of the number of galaxies actually in the field in this range to
those brighter than $L$ multiplied by a weighting factor to account for
the differing volumes sampled. The method generates a cumulative
luminosity function without normalisation.  The differential luminosity
function may be derived by fitting an appropriate model.

The STY method (\cite{STY}) twists this plane around. It examines the
probability that a galaxy observed at a redshift $z_0$ is brighter than
$M$:
\be
\label{eq:STYorig}
P(M,z_0) = { \int_{-\infty}^M \phi(M') D(z_0) f(m') dM' \over \int_{-
\infty}^\infty \phi(M') D(z_0) f(m') dM' }
\ee
where $f(m)$ is the completeness of the survey at apparent magnitude
$m$ and $D(z)$ is the density of galaxies at redshift $z$ divided by the
mean density of galaxies ($\bar{\rho}$). Taking the derivative of this
equation with respect to $M$ yields the probability density for finding
a galaxy with absolute magnitude $M$ in a magnitude-limited survey. 
This probability is directly proportional to the density of galaxies with
that apparent magnitude and inversely proportional to that which
could have been observed at that particular redshift,
\be
\label{eq:STY}
p_k \propto \phi(M_k) \left/
\int_{M_\rmscr{faint}(z_k)}^{M_\rmscr{bright}(z_k)} \phi(M')
dM' \right .
\ee
Here we have replaced the function $f(m')$ with a function that is zero
outside the magnitude limits of the survey and unity within (i.e.\
assuming that the survey is 100\% complete). As the redshift of the 
galaxy is fixed, the discontinuities in $f(m')$ correspond
with the range of absolute magnitudes beyond which no galaxies at this
redshift could have been observed. 

Finally, the Step-Wise Maximum-Likelihood method follows from
equation~\ref{eq:STY} (\cite{EEP}).  Our modification of the SWML
method (SSWML) complements Cho\l oniewski's method, but for a
different statistical model. 

\subsection{Deriving the SSWML Method}
\label{sec:sswml}

The aforementioned clustering-insensitive methods by design probe the
LF as a function of luminosity only. To understand the evolution of 
the LF with redshift, this restriction must be removed. Here we will 
derive two new methods that avoid this restriction. They are based
on, respectively, the STY and SWML methods. Our generalisations of
the locally-valid estimators will be denoted with a prefixed `S' for 
spatial, i.e. SSTY and SSWML respectively.

In the derivation, it becomes apparent that these generalisations 
provide some additional rewards. For example, they provide a 
straightforward prescription to combine various surveys coherently 
and to determine the absolute normalisation.

The derivation of the SSWML method begins with equation~\ref{eq:STYorig} 
of the STY method, modified by two generalisations:
\begin{itemize}
\item $\rho(z,M) \neq \bar{\rho} D(z) \phi(M) $ (i.e.\ luminosity and
density evolution are separately permitted), and
\item $f(m) = \Omega(m)$, where $\Omega(m)$ is the area of sky
sampled at apparent magnitude $m$, accounting for sampling rate and
mean completeness.
\end{itemize}
These two generalisations will allow the determination of the LF 
as a function of redshift (as in \cite{Chol86}) and the use of
many surveys in a single coherent determination of the luminosity
function.

The relevant probability is that of finding a galaxy in the survey
brighter than $M$ and closer than $z$:
\be
P(M,z) = { \int_0^z \int_{-\infty}^M \rho(z', M') 
            \Omega(m') {dV \over dz} dM' dz' \over
            \int_0^\infty \int_{-\infty}^\infty \rho(z', M')             
\Omega(m') {dV \over dz} dM' dz' }
\ee
Again, taking the derivative of this equation yields the probability
density: 
\begin{eqnarray}
\label{eq:genSTY} 
p_k & =&  {\partial^2 P(M,Z) \over \partial M \partial z} \\
    & \propto &
\rho(z_k,M_k) \left/  \int_0^\infty \int_{-\infty}^\infty \rho(z',M')
\Omega(m') {dV \over dz} dM' dz' \right . \nonumber
\end{eqnarray}
where $\Omega(m)$ is the solid angle sampled at the apparent
magnitude $m$, and $m$ is the apparent magnitude corresponding to
$M$ and $z$ considering the distance modulus and effects of the $k$-correction.
Consequently $m$ may best be written $m(z,M;c_k)$ where $c_k$ is the
$k$-correction class of galaxy $k$.  This equation forms the basis of the
SSTY method.

\begin{figure}
\plotone{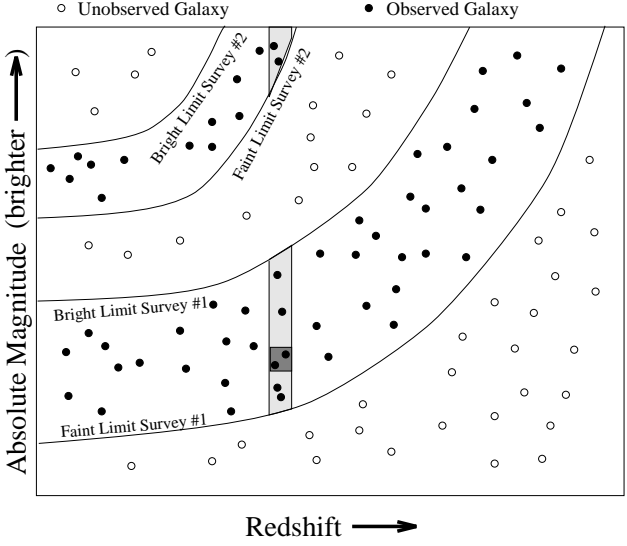}
\caption{Geometric comparison of the STY and SSTY methods. In the STY
method, the probability of observing a galaxy is the ratio of the
number of galaxies at the galaxy's absolute magnitude (the dark-shaded
region) to the total number that could have been observed within the
magnitude limits at the same redshift as the observed galaxy (the
light-shaded region).  The extension to the SSTY technique is
straightforward. The probability of observing a galaxy at a particular
redshift is the ratio of the density of galaxies at the observed
redshift and absolute magnitude (again the dark-shaded region) to the
total number of galaxies that could have been observed at any redshift
(the region enclosed by the magnitude limits).}
\label{fig:sswml}
\end{figure}

Figure~\ref{fig:sswml} compares Equations~\ref{eq:STY}
and~\ref{eq:genSTY} on the absolute magnitude {\it versus} redshift plane. 
The lower double integral is simply the number of galaxies that one
would expect to observe in a combined survey given a LF.  If the 
function $\Omega(m)$ is simply a series of steps (as in 
Figure~\ref{fig:sswml}), there is only one $k$-correction class, and the
trial LF is integrable at a given redshift (like an evolving Schechter
function), the integral may be most efficiently calculated as
\begin{eqnarray}
N_\rmscr{gal}^\rmscr{pred} & = & \int_0^\infty \sum_j \Omega_j
\int_{B_{j,\rmscr{min}}}^{B_{j,\rmscr{max}}} {dV \over dz} dm' dz'
\times \\ \nonumber
 &  & \qquad \rho(z',m'-d_\rmscr{modulus}(z)-k(z)) 
\end{eqnarray}
where the sum is over the subsurveys and the $j$th survey samples
$\Omega_j$ steradians in an apparent magnitude range from
$B_{j,\rmscr{min}}$ to $B_{j,\rmscr{max}}$.  Although the function
$\rho(z,M)$ may not be separable, if the evolution of the LF
is posed as an evolution of separable Schechter parameters,
$\rho(z,M)$ takes the form of an integrable Schechter function 
at all redshifts.

Although now complex, if we reinstate the assumption that the 
density is separable and that we have a single magnitude-limited sample, 
this formula reduces to Equation~\ref{eq:STY}: 
\be
p_k \propto n(z_k) \phi(M_k) \left / \int_0^\infty n(z) {dV \over dz}
\int_0^\infty
\phi(M) \Omega(m) dM' dz' \right .
\label{eq:reduce1}
\ee
We can calculate $\Omega(m)$ for $z_k$
\be
     \Omega(m) = 
\left \{  
\begin{array}{cl}
\Omega & \rmmat{ if } M_\rmscr{faint}(z_k) \le M \le
M_\rmscr{bright}(z_k) \\ 0
&
\rmmat{ otherwise}
\end{array}
\right . .
\label{eq:OmegaSTY}
\ee
Since we seek an estimator for $\phi(M)$ which is assumed constant
everywhere, we can choose to perform the latter integral at the 
redshift of the observed galaxy ($z_k$), yielding  
\begin{eqnarray}
\label{eq:reduce2}
p_k & \propto & n(z_k) \phi(M_k) \left / \phantom{\int_0^\infty}
\right . \\ \nonumber 
	& & \qquad \int_0^\infty n(z) \Omega
{dV \over dz} 
dz \int_{M_\rmscr{faint}(z_k)}^{M_\rmscr{bright}(z_k)} \phi(M) dM 
\end{eqnarray}
which reduces to the STY result.

We now proceed in the spirit of the SWML (\cite{EEP}) method and
assume that $\rho(z,M)$ is defined for a two-dimensional array of steps
in both redshift and luminosity,
\be
\label{eq:rhodef}
\rho(z,M) \equiv \sum_{ij} W(z-z_i,M-M_j) \rho_{ij} \ee where
\be
\label{eq:Wdef}
W(z,M) = \left \{ 
\begin{array}{cl}
1 & \begin{array}{ll}
    \rmmat{ if } & -\Delta z/2 \le z \le \Delta z/2 \rmmat{ and } \\
                 & -\Delta M/2 \le M \le \Delta M/2
    \end{array} \\
0 & \rmmat{ otherwise} 
\end{array}
\right . .
\ee
Substituting this relation into the formula for $p_k$ gives 
\begin{eqnarray}
\label{eq:pk}
p_k & \propto & \sum_{ij} W(z_k-z_i,M_k-M_j) \rho_{ij} \left/
\phantom{\sum_{ij}} \right . \\ \nonumber
 & & \qquad \sum_{ij} \rho_{ij}
\int_{z_i-\Delta z/2}^{z_i+\Delta z/2}
\int_{M_j-\Delta M/2}^{M_j+\Delta M/2} \Omega(m) {dV \over dz} dM
dz  .
\end{eqnarray}
Next we calculate the logarithm of $p_k$
\be
\label{eq:lnpk}
\ln p_k = \sum_{ij} W(z_k-z_i,M_k-M_j) \ln \rho_{ij} - \ln I_k 
\ee
where we have replaced the denominator with $I_k$. The subscript $k$ 
is retained because the value of the integral depends on the $k$-correction 
for the particular galaxy observed through $\Omega(m(z,M;c_k))$.

The likelihood of observing the entire survey is the product of the
likelihoods for each galaxy
\be
\label{eq:Psurvey}
P_\rmscr{survey} = \prod_k p_k,
\ee
and taking the logarithm,
\be
\label{eq:lnPsurvey}
\ln P_\rmscr{survey} = \sum_k \ln p_k.
\ee
Because we would like to maximise the survey likelihood, we examine
the derivative with respect to $\rho_{pq}$ and search for stationary points:
\be
\label{eq:dlnPsurvey}
{d \ln P_\rmscr{survey} \over d \rho_{pq}} = 
     \sum_k {d \ln p_k \over d \rho_{pq}} = 0.
\ee
Differentiating Equation~\ref{eq:lnpk},
\be
\label{eq:dlnpk}
{d \ln p_k \over d \rho_{pq}} =  { W(z_k-z_p,M_k-M_q) \over \rho_{pq}
} - {1 \over I_k}{d I_k \over d \rho_{pq}}.
\ee
Referring to Equation~\ref{eq:pk}, we see that
\be
\label{eq:dIk}
{d I_k \over d \rho_{pq}} = \int_{z_p-\Delta z/2}^{z_p+\Delta z/2}
\int_{M_q-\Delta
M/2}^{M_q+\Delta M/2} \Omega(m) {dV \over dz} dM dz  \ee and
\be
\label{eq:Iksum}
I_k = \sum_{pq} \rho_{pq} {d I_k \over d \rho_{pq}}  
\ee 
Summing over the derivatives (Equation~\ref{eq:dlnpk}) gives 
\be
\label{eq:dlnPsurvey2}
{d \ln P_\rmscr{survey} \over d \rho_{pq}} = \sum_k \left ( {1 \over \rho_{pq}}
 W(z_k-z_p,M_k-M_q) -  {1 \over I_k}{d I_k \over d \rho_{pq}} \right ).
\ee
and setting this to zero yields
\be
\label{eq:rhoiter}
\rho_{pq} =  { \sum_k W(z_k-z_p,M_k-M_q) \over \sum_k \left \{  d I_k
/ d
\rho_{pq} \left / \sum_{ij} \rho_{ij} d I_k / d \rho_{ij}  \right . \right \}
}
\ee
which is familiar in form to the SWML result,
\begin{eqnarray}
\label{eq:SWML}
\phi_q \Delta M & = & \sum_k W(M_k-M_q) \left / \phantom{\sum_k}
\right . \\ \nonumber 
& & \qquad \sum_k \left \{ 
H[M_q-M_{\rmscr{faint}(z_k)}] \left / \phantom{\sum_j} \right . \right
. \\ \nonumber
& & \left . \qquad \sum_j \phi_j \Delta M H[M_j-
M_{\rmscr{faint}(z_k)}] 
\right \}
\end{eqnarray}
since we can identify $I_k$ with the sum over $j$ in the denominator and
$dI/d \rho_{pq}$ with $H[M_q-M_{\rmscr{faint}(z_i)}]$ also in the
denominator.

Calculating the values in the denominator of Equation~\ref{eq:rhoiter}
is less straightforward than in the SWML method, but 
Equations~\ref{eq:dIk}~and~\ref{eq:Iksum} show that this
denominator is a function of the $\rho_{pq}$, the $k$-correction class of
galaxy $k$, the cosmology (through $dV/dz$ and the distance modulus),
and the details of the survey (through $\Omega(m)$). Therefore we can
calculate the values of the $dI_k/d\rho_{pq}$ for each $k$-correction 
class before beginning the iterative solution to Equation~\ref{eq:rhoiter}.

Furthermore, since the integrals $I_k$ are over volume, it is
straightforward to calculate the total number of galaxies expected
given the current values of $\rho_{ij}$,  
\be
\label{eq:ngalpred} n_\rmscr{gal}^\rmscr{pred} = \sum_{ijk} f_k
\rho_{ij} {d I_k \over d \rho_{ij}}  
\ee 
where $f_k$ is the fraction of the galaxies observed in each
$k$-correction class. All $\rho_{pq}$ can then be normalised so that
Equation~\ref{eq:ngalpred} yields the number of catalogue galaxies.
The normalisation can be done either at each iteration by multiplying
the right-hand side of Equation~\ref{eq:rhoiter} by
$n_\rmscr{gal}^\rmscr{observed}/n_\rmscr{gal}^\rmscr{pred}$, or by
multiplying this ratio after the final iteration. The algorithm
converges more quickly (two or three iterations) if the normalisation
is performed at each step.

Errors may be estimated following procedures discussed by \cite{Saun90}:
\begin{eqnarray}
\label{eq:error1}
\sigma(\log \rho_{pq}) & = & (\ln 10)^{-1} \sigma(\ln \rho_{pq})
\nonumber \\
 &
= & (\ln 10)^{-1} \left ( \partial^2 \ln P_\rmscr{survey} \over (\partial
\ln
\rho_{pq})^2 \right )^{-1/2} \\
   & = & (\ln 10)^{-1} \times \nonumber \\ & & \left [ \sum_k \left
\{ W(z_k-
z_p,M_k-M_q) \phantom{ d I_k \over d \rho_{pq}} \right . \right . \\ \nonumber
   &  & \qquad \left . \left . - \left ( \rho_{pq} { d I_k \over d \rho_{pq}} \left /
\sum_{ij} \rho_{ij}
{d I_k \over d \rho_{ij}} \right . \right )^2 \right \} \right ]^{-1/2}.
\nonumber 
\end{eqnarray}  
Additionally, it is straightforward to estimate upper limits for bins 
in which no galaxies are observed. 
\be
\label{eq:upper}
\rho_{pq}^\rmscr{upper} = { 1/2 \over \sum_k \left \{ d I_k / d
\rho_{pq} \left /
\sum_{ij} \rho_{ij} d I_k / d \rho_{ij}.  \right . \right \} } \ee

In this derivation, $\Omega(m)$ includes several important factors.
The simplest way to define $\Omega(m)$ is to calculate the 
area surveyed in each subcatalogue and then multiply this area by the 
sampling rate and completeness. $\Omega(m)$ then represents a series 
of steps with a jump at the bright and faint limits of each 
subcatalogue. However, we can generalise this in two ways. Firstly, 
since we explicitly calculated $d I_k / d \rho_{ij}$ for each galaxy 
type, we can introduce completeness as a function of $k$-correction 
type by defining various $\Omega_k(m)$. Secondly, we can calculate 
the completeness rate within a subcatalogue as a function of apparent 
magnitude and account for this in obtaining $\Omega(m)$ or $\Omega_k(m)$.
These two generalisations make this technique an extremely versatile 
tool in analysing galaxy catalogues.

\section{Tests and Comparisons}
\label{sec:lfesttest}

To test the \vm\ and SSWML methods in a large sample like the
composite DARS, AUTOFIB and LDSS-1/2 surveys, we generated a random
galaxy catalogue from a Schechter function (\cite{Love92}).  In the
first test 1800 galaxies were selected in a standard cosmology, 300
from each of the following magnitude ranges: 11.0--17.3, 17.0--19.7,
19.7--20.5, 20.3--20.8, 20.8--22.5, and 22.5--24.0. All the galaxies
were assigned zero $k$-correction and the density of galaxies was
doubled beyond a redshift of 0.2 to crudely simulate density
evolution.

\begin{figure}
\plotone{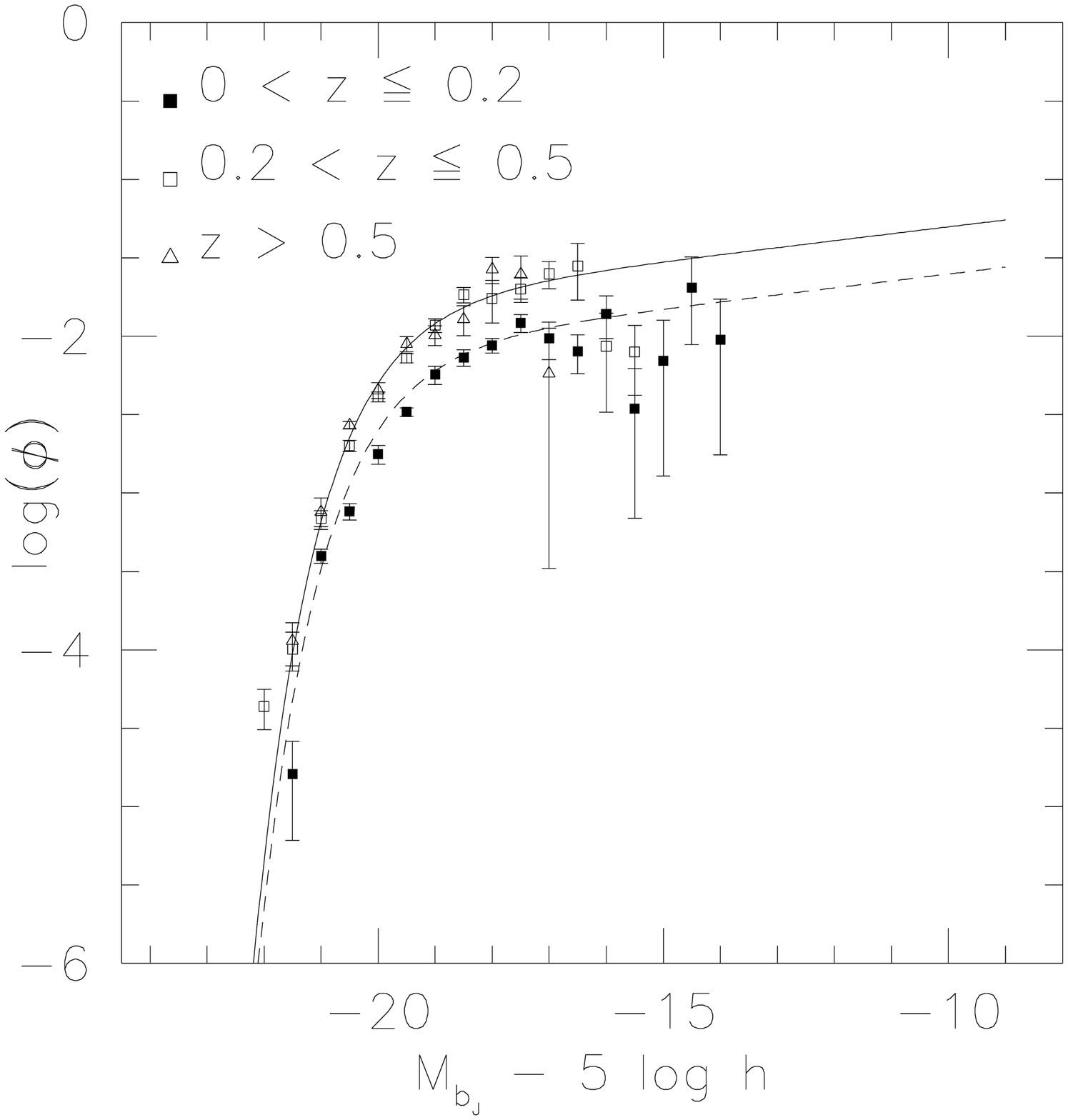}\\
\plotone{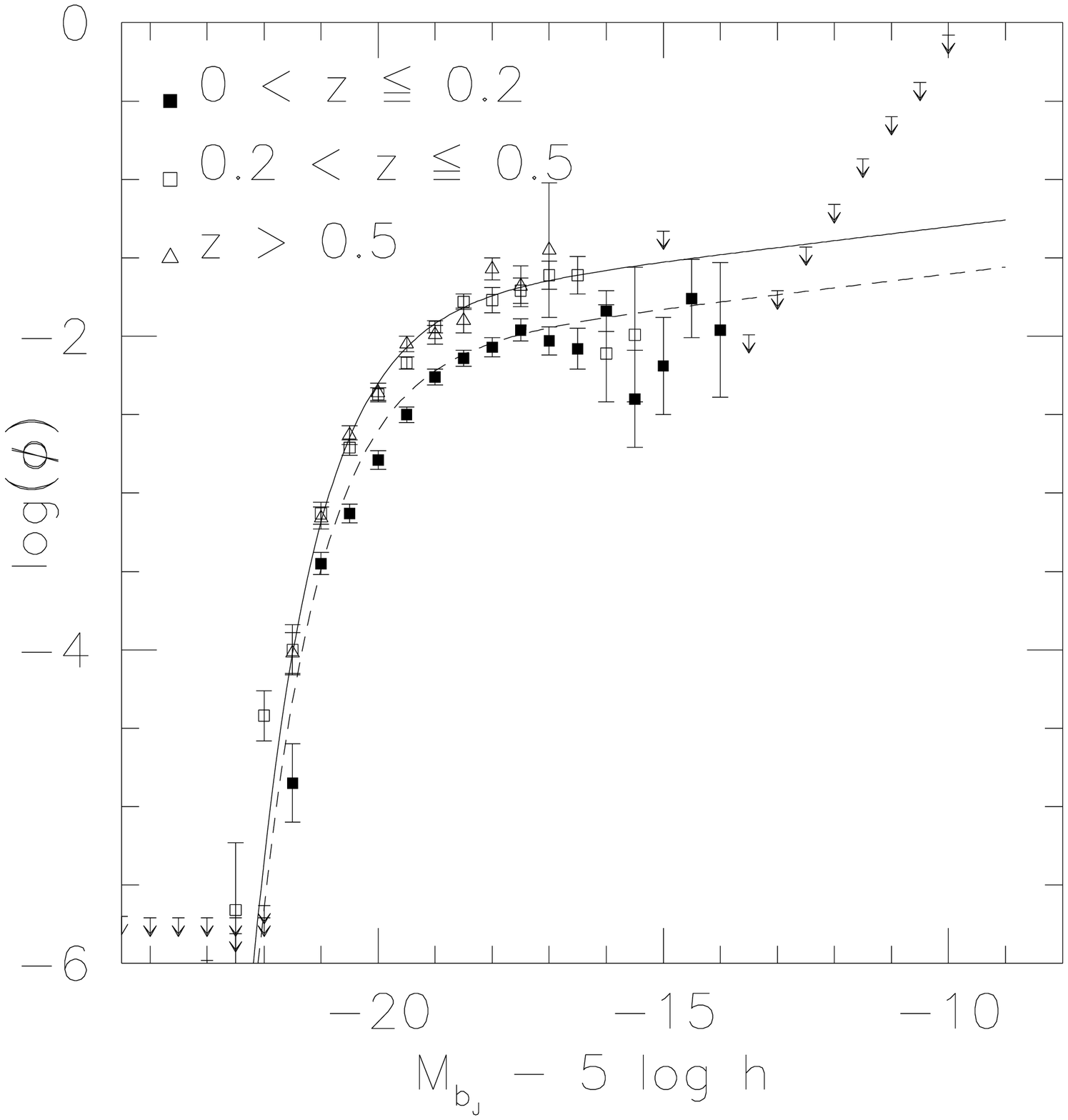}
\caption{Test catalogue of 1800 galaxies.  The upper panel depicts the
results of the \vm\ analysis on the random galaxy catalogue and the
lower panel shows the results of the SSWML method. In each panel the
lower curve is the nearby ($z<0.2$) luminosity function, and the upper
is the distant one. The errorbars for the \vm\ method are generated
using a bootstrap technique, while the errors and upper limits for the
SSWML are determined as described in the text.}
\label{fig:test300}
\end{figure}

Figure~\ref{fig:test300} illustrates the results of this
simulation. It can be seen that the two methods generate nearly
identical LFs, although the SSWML method produces smoother
results.  The bootstrap errors of the \vm\ method agree remarkably
well with errors derived for the maximum likelihood technique.
Although in principle the SSWML technique has the additional advantage
of determining upper limits which further constrain the evolution of
the faint-end slope, these upper limits increase with a slope of
$\alpha=-2$ and so in practice are not very interesting much fainter
than the observed points.

To test whether either of the techniques are biased and to find an
external error measurement, we generated an ensemple of ten catalogues
derived from the same evolving LF as depicted in
Figure~\ref{fig:test300}.  Each catalogue contained 1,800 objects
sampled as in the previous test.  We found that neither method is
systematically biased relative to the input luminosity function and that
the internal error estimates satisfactorily approximate the scatter of 
the luminosity functions derived for the ensemble.

\subsection{Sensitivity to Clustering}

To test the sensitivity of the algorithms to clustering, three
additional galaxy catalogues were generated with an overdensity of
galaxies at a redshift of 0.05 with a dispersion in redshift of 0.003
in one of the fields in the 17.0--19.7 survey. The `cluster' LFs were
assumed identical to that in the field, but were normalised such that
a set fraction of the 300 galaxies in the 17.0--19.7 magnitude
range were in the cluster. The three surveys had clustered fractions
of 45\%, 65\% and 85\%.

\begin{figure}
\plotone{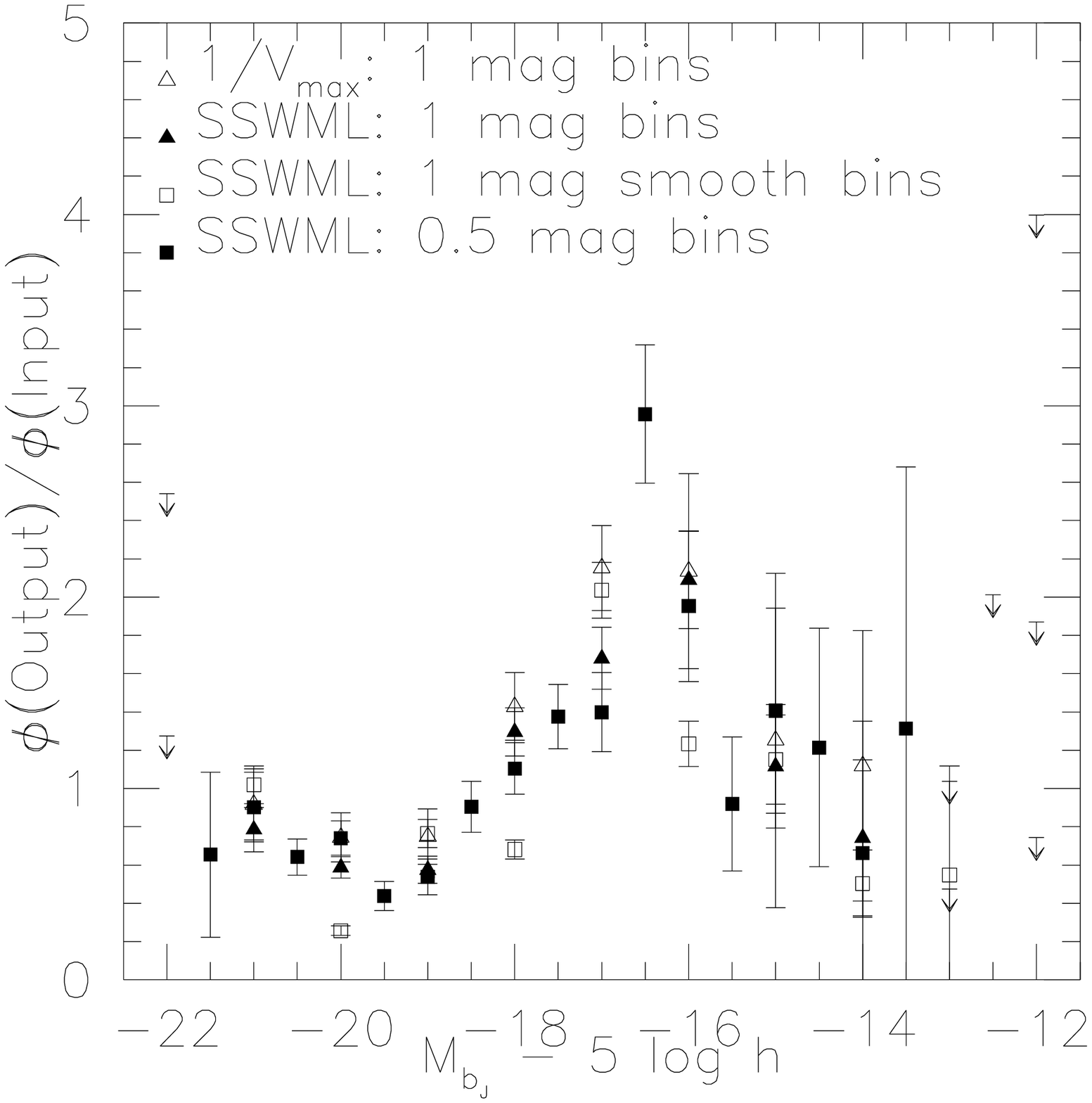}
\caption{Clustered test catalogue of 1800 galaxies. The luminosity
function of the clustered test catalogue was determined using a
variety of methods (see text).}
\label{fig:test300c}
\end{figure}

Figure~\ref{fig:test300c} shows the results of this simulation for the
most populous cluster for a variety of bin widths.  The \vm\ and the
narrowly-binned SSWML methods both over-predict the faint end of the
LF. However, as the bin width of the SSWML method is increased, the
over-prediction decreases. The `smooth binning' technique uses two
SSWML calculations with bins of two magnitudes, shifted by one
magnitude relative to each other. This achieves the closest results to
the input LF.

\begin{figure*}
\plottwow{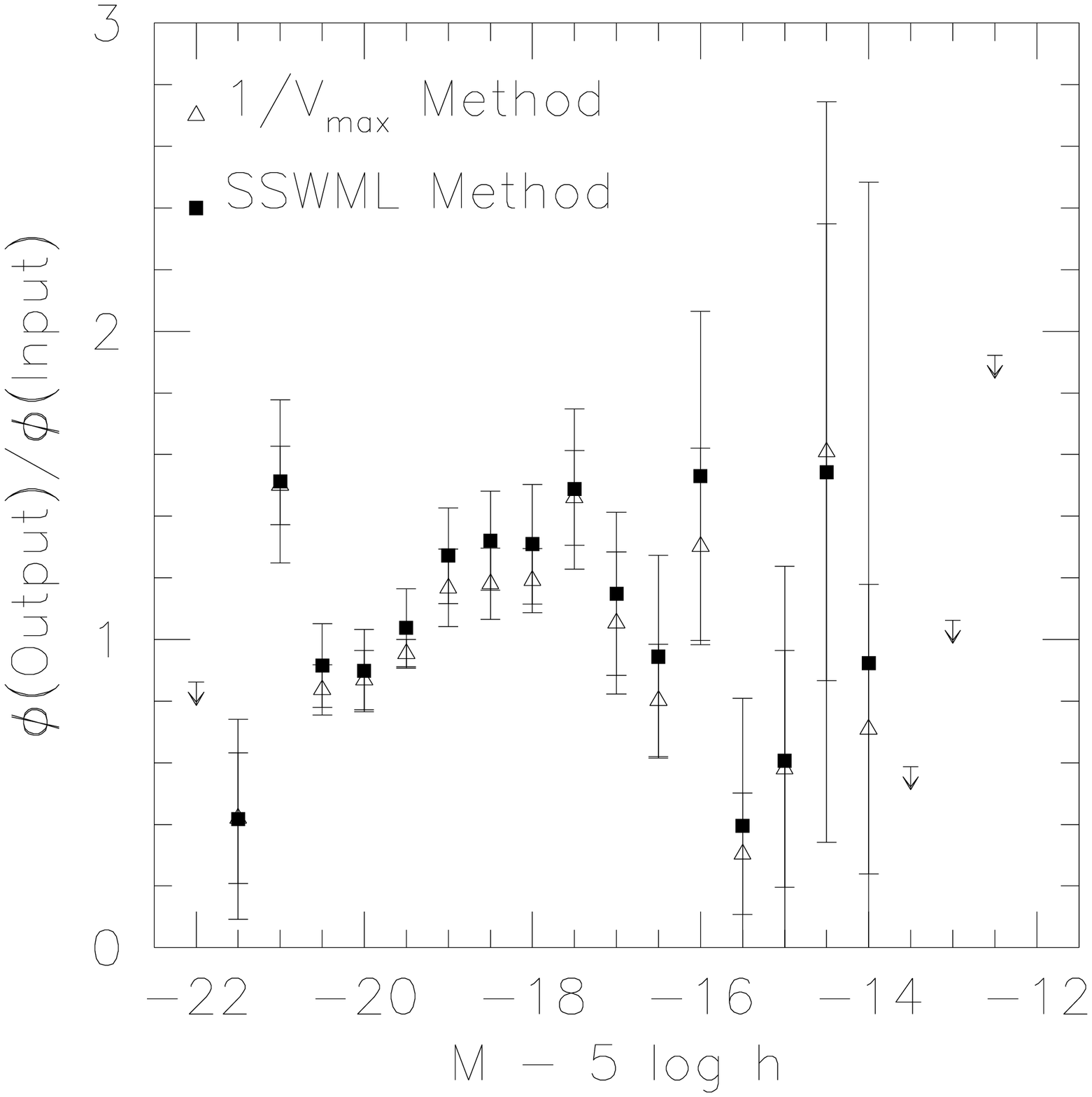}{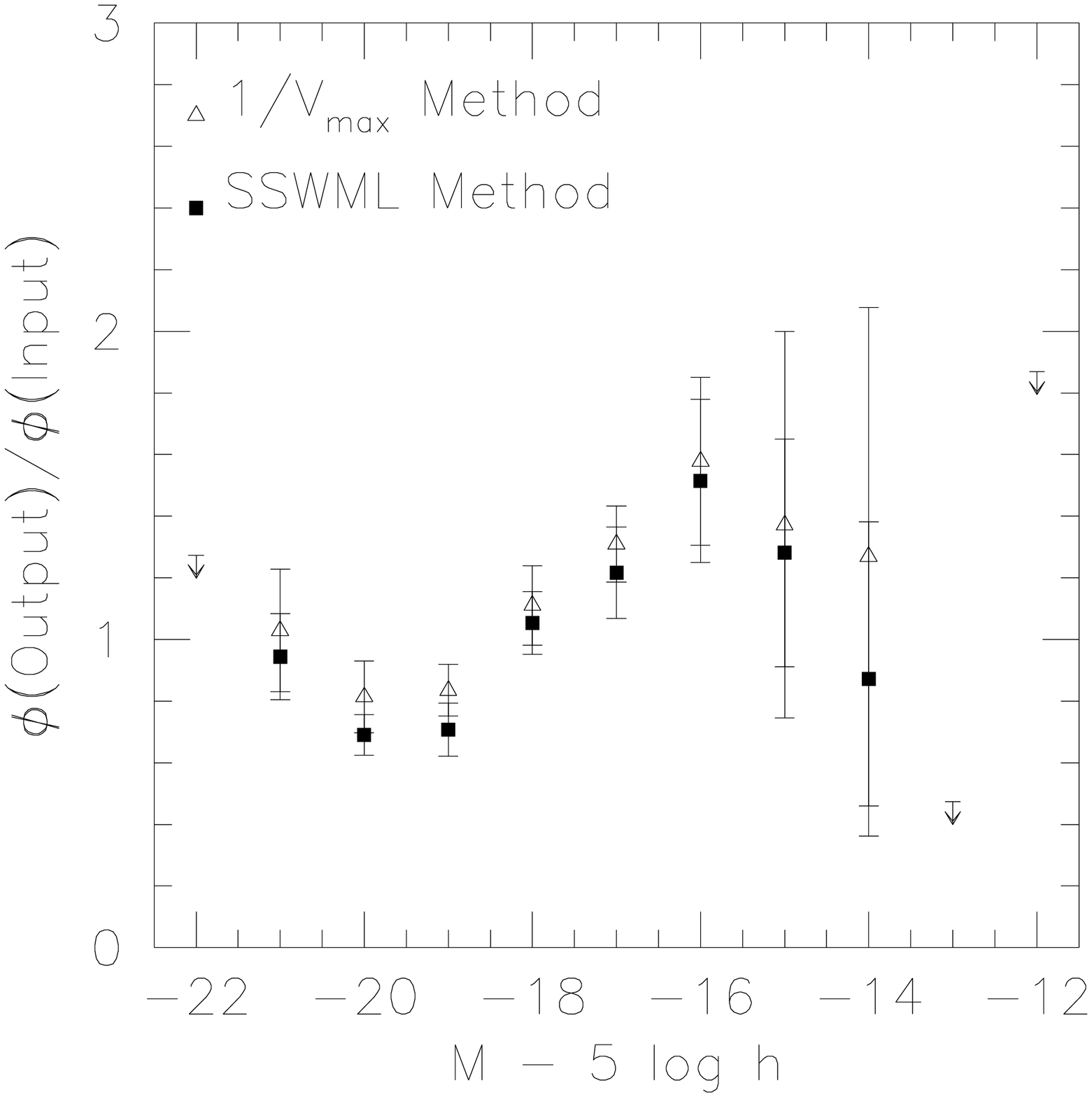}\\
\plottwow{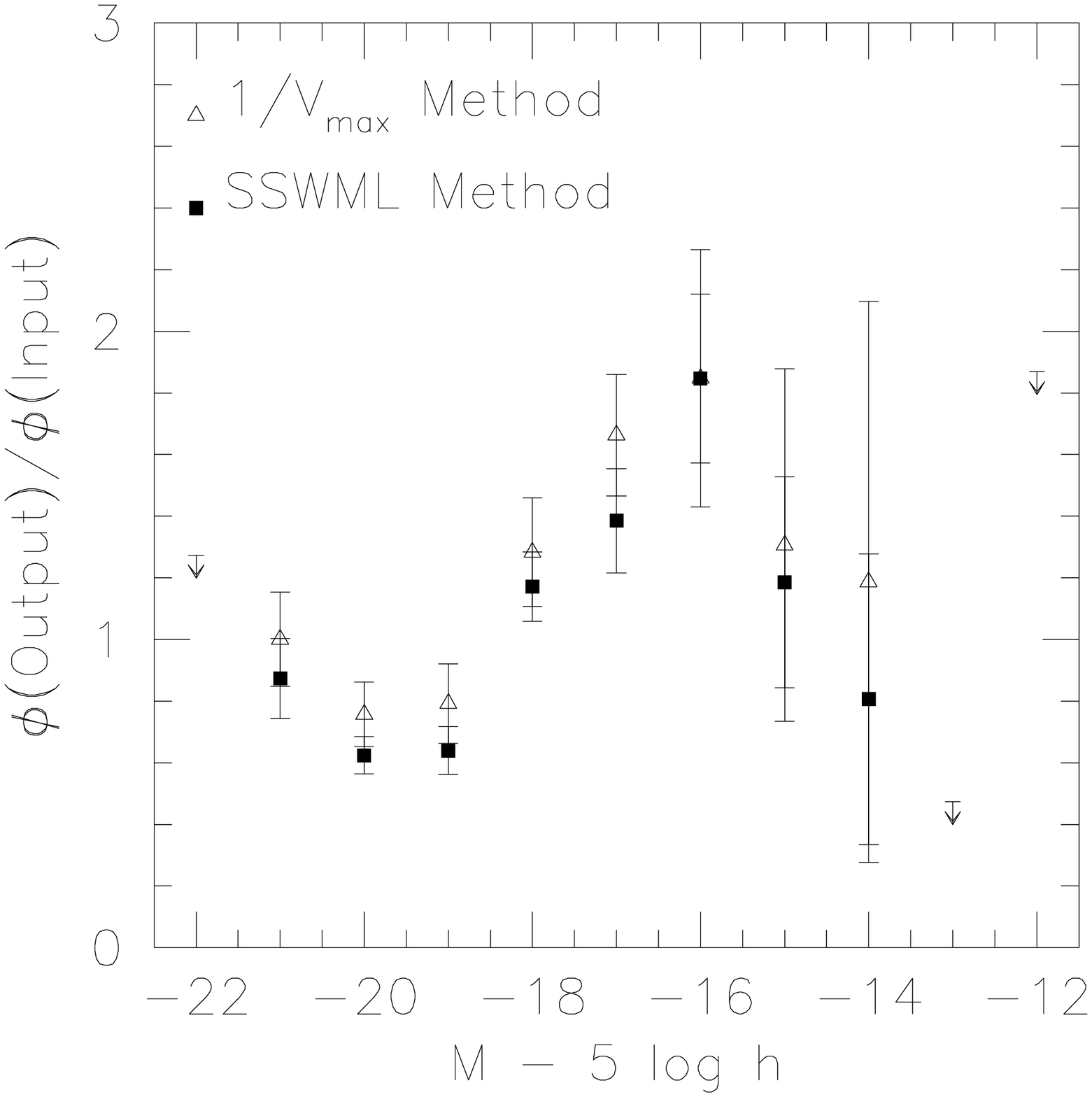}{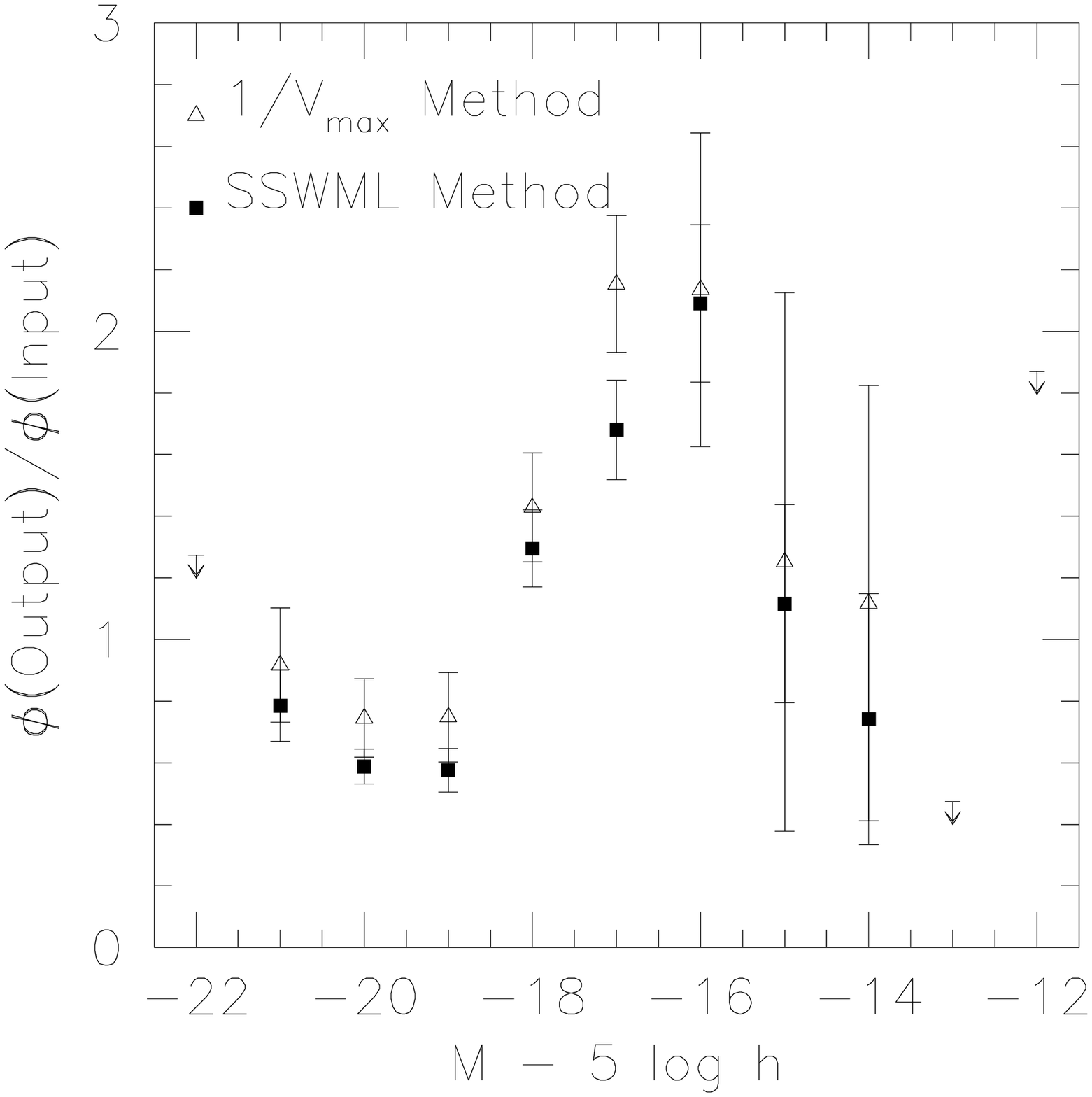}
\caption{Luminosity functions ($0.02 \leq z < 0.15$) as a function of
clustering. The top-left panel has no clustered galaxies. In the
top-right panel, 45\% of the galaxies in the B=17--19.7 subcatalogue
are in a cluster at $z$=0.05. In the lower panels 65\% (left) and 85\%
(right) of these galaxies are clustered.}
\label{fig:testcomp}
\end{figure*}

Figure~\ref{fig:testcomp} compares the LFs derives from the 3
clustered catalogues and the unclustered catalogue. Both the SSWML and
\vm\ methods derive successively steeper LFs as the strength of the
clustering is increased, whereas the SSWML esimator is clearly less
affected.  In summary, these tests reveal the superiority of the SSWML
method in surveys affected by clustering.

\subsection{The Relation between the Methods}

Figure~\ref{fig:test300c} reveals a interesting correspondence between
the \vm\ and the maximum-likelihood technique.  It appears that the
 smaller the SSWML bins, the closer it approximates the \vm\
method. Looking back to Equation~\ref{eq:rhoiter}, the sum over the
$k$ objects in the survey in the numerator may be replaced with
$n_{pq}$, the number of objects in the appropriate bin. Furthermore, 
consider the case where there is only one type of galaxy. This is not
a restrictive assumption, as one could split the surveyed galaxies by
type, calculate the LF of each type independently prior to addition.
These two approaches result in
\be
\rho_{pq} = {n_{pq} \over n_\rmscr{gal}^\rmscr{observed}} { \sum_{ij}
\rho_{ij} d I
/ d \rho_{ij} \over d I / d \rho_{pq}}
\ee
where the sum over $k$ has been replaced by a multiplication. Referring
to Equation~\ref{eq:ngalpred}, the sum in the numerator may be
replaced with $n_\rmscr{gal}^\rmscr{predicted}$, yielding 
\be
\rho_{pq} = {n_{pq} \over n_\rmscr{gal}^\rmscr{observed}} {
n_\rmscr{gal}^\rmscr{pred} \over d I / d \rho_{pq}}. 
\ee
It appears that the iterative process is not required in this
simplified case. The normalisation of $\rho_{pq}$ must be determined
by multiplying by
$n_\rmscr{gal}^\rmscr{observed}/n_\rmscr{gal}^\rmscr{pred}$, yielding
\be
\label{eq:rhopq2}
\rho_{pq} = {n_{pq} \over d I / d \rho_{pq}}
\ee
The final connection is Equation~\ref{eq:dIk} in the limit when $\Delta
M$ is small, such that $\Omega(m)$ can be assumed to be constant
across $\Delta M$. As a survey has magnitude limits this can only be
an approximation. In this way, the inner integral may be approximated
by a product:
\begin{eqnarray}
\label{eq:dI2}
{d I  \over d \rho_{pq}} & \approx & \int_{z_p-\Delta z/2}^{z_p+\Delta z/2}
\Omega(m(z,M_q)) \Delta M {dV \over dz} dz \\ \nonumber
 & \approx & V_{pq,\rmscr{max}} \Delta M.
\end{eqnarray}
Substituting Equation~\ref{eq:dI2} into Equation~\ref{eq:rhopq2} yields
the familiar \vm\ equation,
\be
\rho(z_p,M_q) = \rho_{pq} \approx {n_{pq} \over V_{pq,\rmscr{max}}
\Delta M } \ee
and we have come full circle --- in the limit of small binwidth the
SSWML and \vm\ methods are identical.

\begin{figure}
\plotone{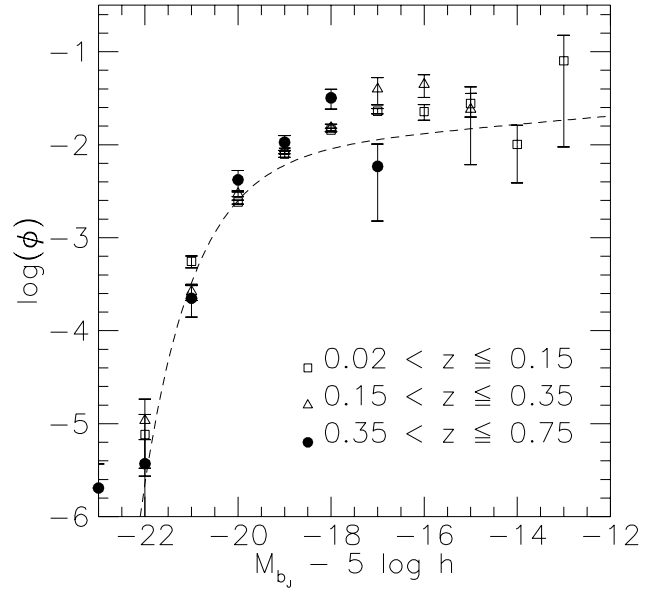} \\
\plotone{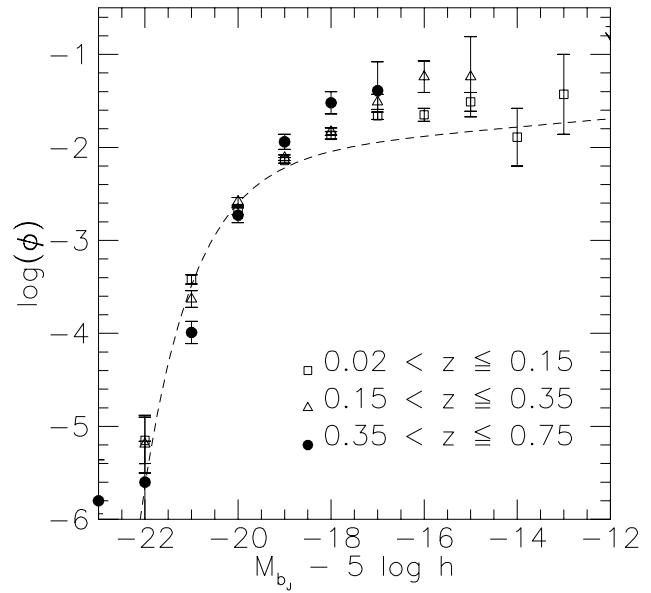} 
\caption{Evolution of the luminosity function estimated by the \vm\
method (top) and the SSWML method (bottom). The luminosity function at
several redshifts is depicted by the symbols in the legend with
1$\sigma$ errorbars.  The long-dashed curve is the best-fit Schechter
function determined by Loveday et~al.\ (1992) without correction for
Malmquist bias.  All the luminosity function figures that follow have
a similar format.}
\label{fig:elfvm}
\end{figure}

Finally, we compare the results of the \vm\ and SSWML methods when
applied to our combined redshift catalogue. Figure~\ref{fig:elfvm}
compares the redshift evolution of the luminosity function obtained
with the \vm\ method (equivalent to Figure~10 of Paper~I, though with
broader bins) and that determined using the SSWML method---the
agreement is extremely good, as implied by the above discussion. In
the rest of this paper we therefore show only the LFs estimated using
the SSWML method, although we have checked and found agreement with
the \vm\ method in every case.

\section{The Evolution of the Luminosity Function by Spectral Type}
\label{sec:typelf}

In Paper I (and Figure~\ref{fig:elfvm}) we demonstrated evolution in
the shape of the overall LF since $z$$\sim$0.75.  We present these
results again here using both the traditional \vm\ technique and the
clustering-insensitive SSWML method. A comparison of these figures
demonstrates that, as we argued in Paper~I, the Autofib survey is
not grossly affected by clustering.  However, the SSWML techinique
yields a slightly different and a considerably more robust estimate
of the galaxy
luminosity function.

We also showed in Paper~I how the steepening of the faint end slope
with look-back time is caused primarily by the population of galaxies
with strong [OII] emission, which effectively trace the star-forming
component of the field galaxy population.  The techniques described in
this paper allow us to directly examine the evolution of the galaxy LF
as a function of spectral type.

\subsection{Misclassification and $K$-correction Error}

\begin{figure}
\plotone{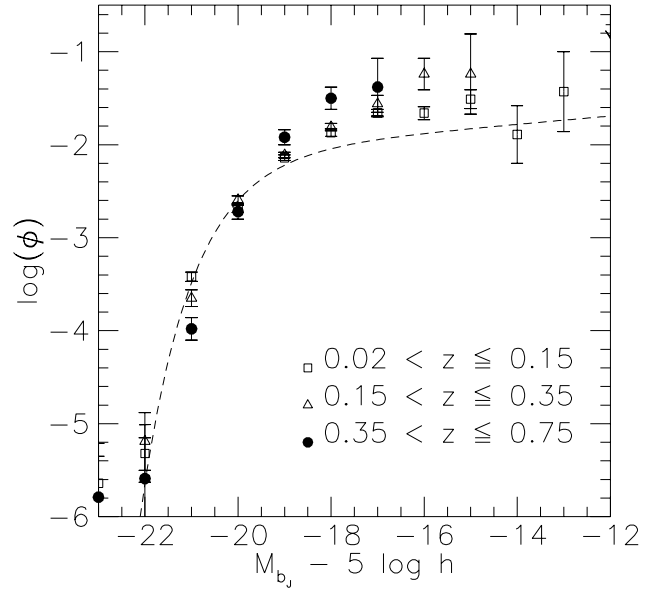}
\caption{Evolution of the luminosity function with 20\% random
re-classifications by one spectral class.} 
\label{fig:elfrand}
\end{figure}

The analysis in \S\ref{sec:kc} shows that the cross-correlation
technique may misclassify approximately one-fifth of the spectra in
the survey by one spectral class. Therefore before proceeding we must
check that such misclassification does not seriously affect our
results. To address this, 20\% of the galaxies were randomly reclassified
by one class blueward or redward, the sampling volumes and absolute
magnitudes re-calculated, and the luminosity functions
re-determined. Figure~\ref{fig:elfrand} presents the results of this
test. These luminosity functions agree with those presented in
Figure~\ref{fig:elfvm} to within the estimated errors, showing that
the luminosity functions derived are not sensitive to the level of
misclassification present in our sample.

\begin{figure}
\plotone{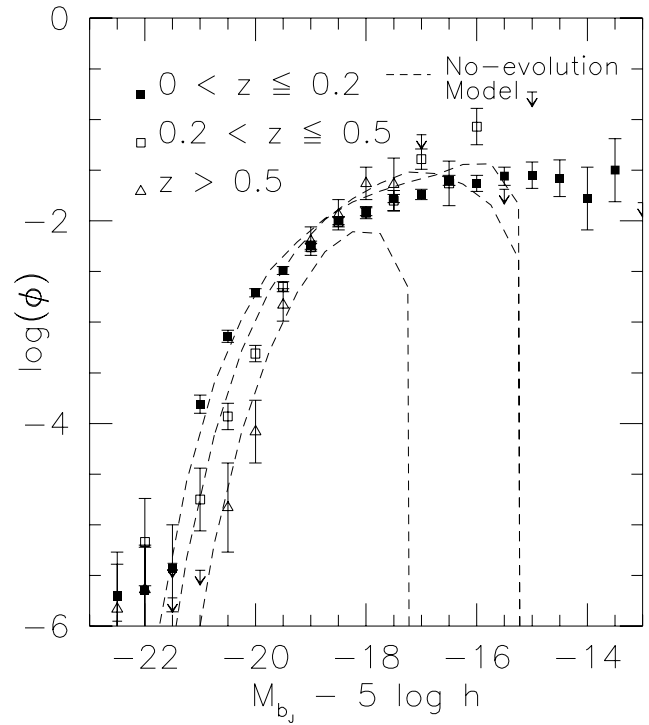}
\caption{Evolution of the luminosity functions in the observer's frame.}
\label{fig:elfobsf}
\end{figure}

To test the sensitivity to the $k$-corrections themselves, we
calculate the evolving luminosity function without converting the
observed magnitudes to the rest-frames of the galaxies.  The evolving
observer's frame luminosity function is calculated using the SSWML
method (Figure~\ref{fig:elfobsf}). Here, the $k$-corrections
determined for the surveyed galaxies play no role in the derivation of
the luminosity functions.  For comparison, we calculate how the
observed local distribution of galaxies would appear at higher
redshifts if they are observed in the $B$-band and binned by their
observed $B$-band magnitude, i.e.\ the `no-evolution' prediction.  To
perform this comparison we assume a specific local mix of galaxies
with varying $k$-corrections.  As the galaxies suffer larger and
larger $k$-corrections, the bright-end cutoff luminosity becomes
markedly fainter with redshift in both the observations and the model.
However, the model predicts fewer faint galaxies than are observed.
This effect is analogous to the increase in the faint-end slope in the
rest-frame luminosity functions. The number of faint galaxies must
therefore increase with look-back time, independent of possible errors
in the $k$-corrections that we have derived.

\subsection{Characterising the Evolution}

For convenience, we divide the sample into three broad spectral types:
E/S0s, early spirals (Sab/Sbc) and late spirals (Scd/Sdm/starburst) in
accordance with the discussion in $\S$2. For each of these subsamples we
compare the co-added spectra at low and high redshift. The spectra were
coadded in the rest-frame with normalisations which are allowed to vary
in order to minimise the mean dispersion between the various spectra. 
Since each spectrum covers a slightly different wavelength range, this is 
not exactly equivalent to normalising according to the total flux in 
each spectrum, but the difference only marginally affects the coadded 
spectra in the range of interest. Because of uncertainties in fluxing 
these data drawn from such diverse sources, we have made no attempt 
to flux-calibrate these data. Rather, they are intended to illustrate 
possible evolution of the {\it mean spectral features} of the galaxies.

We also determine the LF as a function of redshift for each spectral type
by the SSWML method, using a coarser sampling than for the combined
sample due to the smaller numbers within each spectral class. As a
different approach to characterising the LF evolution of each class,
we fit a model for the evolution of the Schechter parameters with
redshift using the generalised STY (SSTY) method derived in
Equation~\ref{eq:genSTY}. This is done by maximising the likelihood of
observing each subsample assuming the LF evolution takes the following
form:
\begin{eqnarray}
\phi^*(z) & = & \phi^*_0 (1 + z)^{\gamma_\phi} \nonumber \\ 
L^*(z)    & = & L^*_0 (1+z)^{\gamma_L} \\
\alpha(z) & = & \alpha_0 + \gamma_\alpha z. \nonumber
\label{eq:evoldef}
\end{eqnarray}
In the above, both the normalisation and characteristic luminosity
evolve as a power of cosmic time ($\phi^* \propto t^{-3\gamma_\phi/2}$
and $L^* \propto t^{-3\gamma_L/2}$), while the faint-end slope evolves
linearly with redshift. As with a non-evolving Schechter function, the
parameters are highly correlated.  The density of galaxies of a given
class is given by:
\begin{eqnarray}
\phi(L,z) &=& \phi^*_0 (1+z)^{\gamma_\phi - \gamma_L (\alpha_0 +
\gamma_\alpha z)} 
     \left ( {L \over L^*_0} \right ) ^ {\alpha_0 + \gamma_\alpha z} \times
\\ \nonumber
 & & \qquad \exp \left ( {-L \over L^*_0 (1+z)^{\gamma_L} } \right ).
\end{eqnarray}
For example, if the faint-end slope does not change with redshift
($\gamma_\alpha$=0), then $\phi^*_0$ and the product $\alpha_0 L^*_0$
jointly determine the density of faint galaxies as a function of
redshift. The correlation of the six parameters makes the errors
difficult to interpret. However, although the trends in individual 
parameters are uncertain, those which determine the overall LF shape are
much better determined (at least within the evolutionary formalism
adopted). Although the best-fit parameters maximise the likelihood of
observing the survey, several sets of these parameters will produce
nearly identical trends within the absolute magnitude and redshift
range probed.

Furthermore, it is essential to note that the given SSTY model is
restricted to monotonic evolution of the parameters with redshift.  It
will find the trend that the bulk of the galaxies of a given spectral
type follow.  Since the catalogue is numerically dominated by galaxies
at moderate redshift, the SSTY model is most strongly sensitive to the
recent evolution of the population.
   
\subsection{Early-Type Galaxies}
\label{sec:es0lfev}

Galaxies with early-type (E/S0-like) spectral classifications represent
20\% of the combined survey (331/1603 classified galaxies) and
contribute approximately one-tenth of the volume density of
intrinsically faint galaxies and one-half of the galaxies with $L\sim
L^*$.

\begin{figure}
\plotsmlandmn{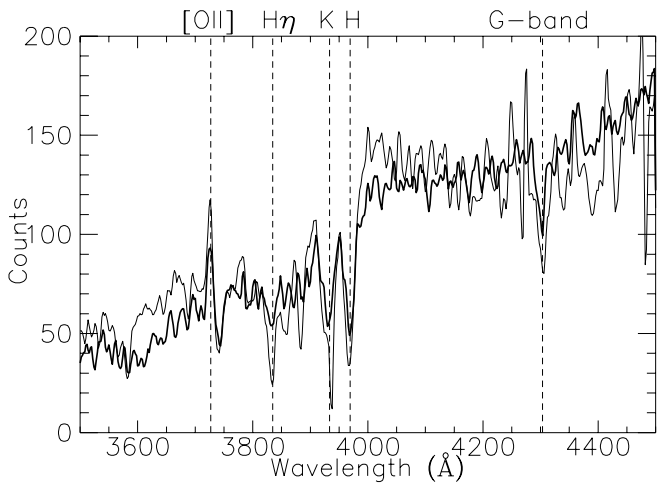}
\caption{Coadded spectra for early-type galaxies.  All the galaxies with
E/S0 spectral classifications have been coadded in two groups. The
bold curve is the coaddition of all early-types with redshifts
$z$$<$0.2. The light curve is for those with 0.2$<$$z$$<$0.5.  Both
spectra have been smoothed on a scale of 10\AA.}
\label{fig:coaddell}
\end{figure}

Figure~\ref{fig:coaddell} compares the mean spectra of the E/S0
galaxies with $0<z\le 0.2$ with those $0.2<z\le 0.5$. At first sight
the two spectra appear very similar to each other and also to the E/S0 
spectra given in the Kennicutt atlas (\cite{Kenn92a}). The most striking
differences concern the 4000 \AA\ break and [O II] emission which 
{\it both} appear to strengthen with redshift. The former effect is surprising
unless there is a selection bias in favour of detecting galaxies
with strong features and so cannot be considered a reliable result given 
the absence of a flux calibration. The presence of the [OII] line in the 
coadded spectra is more interesting, although it could readily arise 
from the presence of a few later-type systems with red spectra, as evidenced
in the discussion of the precision of our classification system in $\S$2.5. 

The important point here is the absence of evidence for recent star 
formation in this predominantly spectrally-inactive class (e.g. 
a prominent $H\delta$ line, \cite{Zabl96}). This indicates 
that the bulk of the redder sources in our survey have not 
recently undergone any star formation on timescales of 1-2 Gyr and 
thus are passively evolving old systems over the redshift ranges probed.
 
\begin{figure}
\plotone{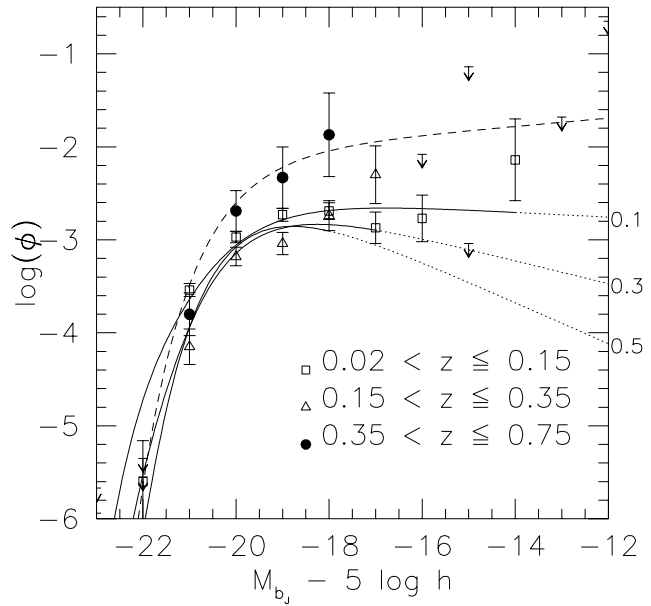}
\caption{The SSWML luminosity functions of E/S0 galaxies in three
redshift ranges are shown by the points. The results of the SSTY LF fits
for $z$=0.1, 0.3 and 0.5 are superimposed.  The best-fitting
monotonically evolving model (found by the SSTY method) underpredicts
the number of elliptical galaxies at high redshift relative to the less
constrained SSWML results. The Loveday et~al.\ (1992) LF (dashed curve)
is shown for reference.  In this and the subsequent LF figures, we have
extrapolated the SSTY LF fits faintward of the observed data to
illustrate the trends more clearly. Extrapolations are defined via a
dotted line.}
\label{fig:elfell}
\end{figure}

Figure~\ref{fig:elfell} shows the evolution of the associated LF determined
using the SSWML method in three redshift ranges, together with the LFs
resulting from SSTY fits for the parameters of the model described
above. Neither method finds any convincing evolution of the E/S0 LF
out to $z$=0.35, while the small sample size at higher redshifts
produces large uncertainties and the two methods suggest
some evolution, but in opposite senses. The overall conclusion, from
both the co-added spectra and the LF analysis, is that there has been
little significant change in the properties of the early-type field 
galaxy population to at least $z$$\sim$0.5.
 
\subsection{Early-Type Spiral Galaxies}
\label{sec:earlyslfev}

\begin{figure}
\plotsmlandmn{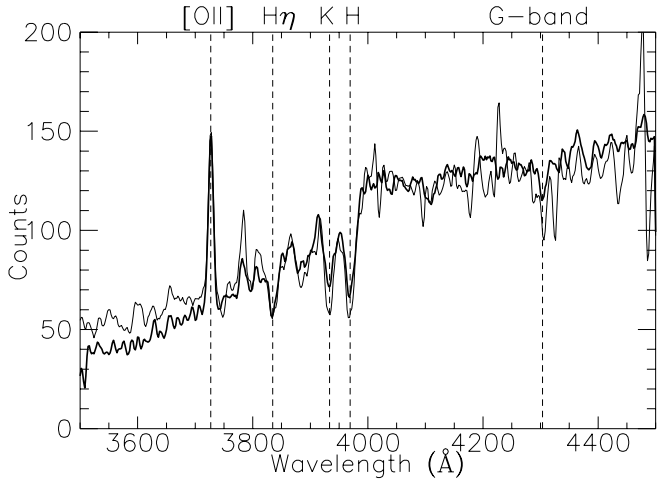}
\caption{Coadded spectra for early-type spiral galaxies. The bold
curve is the coaddition of all early spirals with redshifts $z$$<$0.2,
while the light curve is those with 0.2$<$$z$$<$0.5.  Both spectra
have been smoothed on a scale of 10\AA.}
\label{fig:coaddsab}
\end{figure}

Early-type spirals (Sab/Sbc) comprise 40\% of the
spectrally-classified galaxies (643/1603) and contribute about
one-half of the local density of galaxies for luminosities probed by
the combined survey. Our analysis proceeds as for the early-type
galaxies. Figure~\ref{fig:coaddsab} compares the coadded spectrum for
the early spirals with $z$$<$0.2 and 0.2$<$$z$$<$0.5.  An effect
similar to that found in the early-types is also evident here, with
the absorption lines in the high-redshift spectrum appearing somewhat
stronger than in the low-redshift spectrum. The difference between the 
blue continuum slopes of the two spectra is also in the same direction
as in Figure~\ref{fig:coaddell} and presumably results from the rapid 
decrease in the spectrograph response blueward of 3600\AA\ which
preferentially affects the spectra of the low-redshift galaxies. 
Again, the principle result is the remarkable similarity of the diagnostic
spectral features over this redshift range. We find no significant 
change in the relative contributions of the Balmer absorption lines or the 
[OII] 3727\AA\ emission line between the low- and high-redshift 
spectra (the median EWs of [OII] are similar). 

\begin{figure}
\plotone{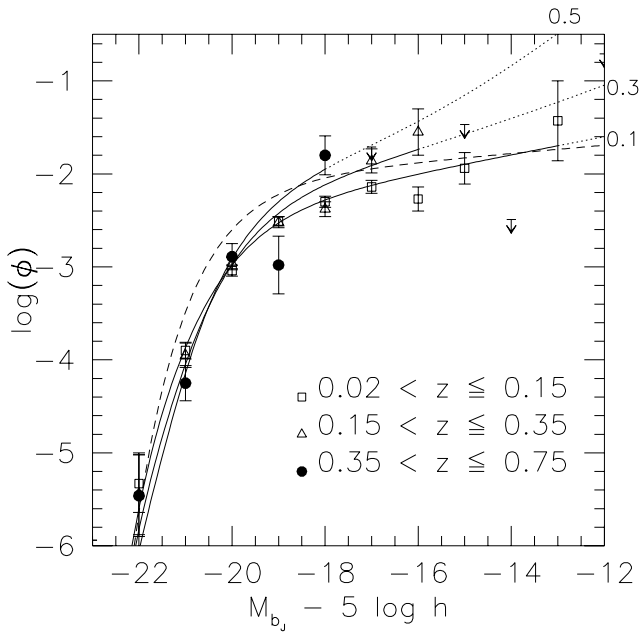}
\caption{SSWML luminosity function of early spiral galaxies. The
results of the generalised STY fitting for the early-type spirals are
superimposed for $z$=0.1, 0.3 and 0.5.}
\label{fig:elfsab}
\end{figure}

However, as Figure~\ref{fig:elfsab} shows, there is some modest
evolution of the early spirals' LF with redshift. Although there is
little or no change in the number of objects with $L$$\gs$$L^*$, the
faint end of the LF steepens as we go to higher redshift. The absence
of post-starburst features in the high redshift coadded spectrum 
indicates the bulk of the population is undergoing constant or
smoothly-declining star formation on timescales of a few Gyr. 

\subsection{Late-Type Spiral Galaxies}
\label{sec:lateslfev}

As shown in Paper I, late-type galaxies appear to be the principal 
source of evolution in the shape of the overall galaxy LF.
In Paper I, the selection method that produced this conclusion was 
based only on the EW of the [O II] emission line. In fact this 
selection correlates reasonably closely with the late-type spiral 
class discussed here and thus we can anticipate similar results. 

\begin{figure}
\plotsmlandmn{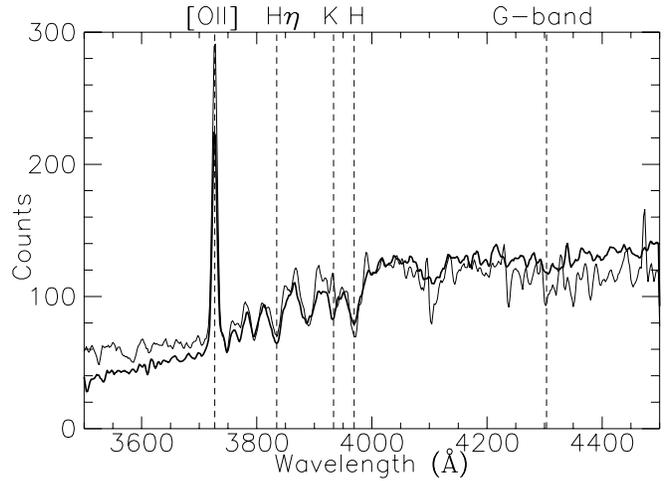}
\caption{Coadded spectra for late-type spiral galaxies. The bold curve
is the coaddition of all late spirals with redshifts $z$$<$0.2, while
the light curve is those with 0.2$<$$z$$<$0.5.  Both spectra have been
smoothed on a scale of 10\AA.}
\label{fig:coaddsdm}
\end{figure}

Importantly, the coadded spectra for galaxies of this class
(Figure~\ref{fig:coaddsdm}) shows a marked change in the nature of the 
star formation present. As originally discussed by \jcite{BES}, the 
higher redshift spectrum shows much deeper $H\delta$ absorption and a
possible reversal of the Ca II line ratio suggesting 
the presence of a stronger $H\epsilon$; these lines are signatures of 
more recent ($\simeq$1 Gyr) star formation. Coupled with this is a
strong increase in the strength of the [O II] emission. These diagnostics
suggest such galaxies are being seen at an atypical stage in their
star formation history. 

\begin{figure}
\plotone{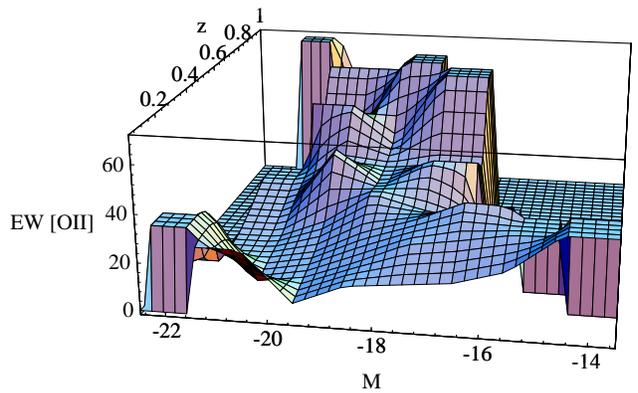}
\caption{Evolution of the median [OII] EW for late-type spirals.  In
regions of the $M_B-z$ where no galaxies were observed, a median EW of
$0\AA$ is plotted.
}
\label{fig:ewsdm}
\end{figure}

Figure~\ref{fig:ewsdm} reveals the [O II] evolution more clearly: at
fixed luminosity there is an increase in the median [OII] EW with
redshift. This effect is superimposed upon the well-known tendency for
fainter objects to have stronger emission. Analysing objects in a narrow
range of spectral type minimises the bias arising from the fact that
fainter samples become more dominated by such galaxies because of the
$k$-correction (\cite{Koo92}). Taking [OII] EW as an indicator of
star-formation rate, we can therefore conclude that in late-type spirals
the star-formation rate is higher in lower luminosity objects at fixed
redshift and higher in higher redshift objects at fixed luminosity.
Alternatively, a given star-formation rate is found at higher redshift
in higher luminosity objects. The spectra indicate the nature of
star formation changes also, becoming more `burst-like' in the distant
samples.

The evolution of the late-type spirals dominates the more modest
changes seen in the LFs of the other types.
Their density and faint-end slope both increase with redshift, while 
the cutoff luminosity remains fixed. Consequently they contribute an 
increasing fraction at all luminosities, progressively dominating the 
population at brighter limits and by $z$$\sim$0.5 becoming the most 
common galaxy type at all luminosities below $L^*$.

\begin{figure}
\plotone{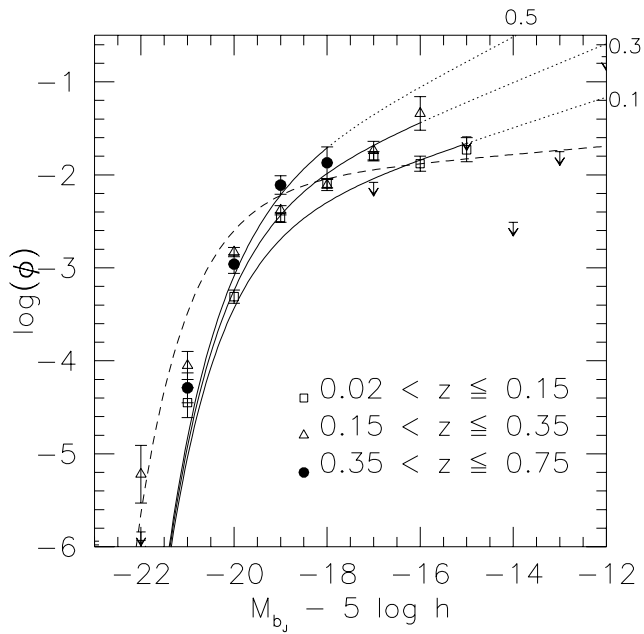}
\caption{SSWML luminosity function of late spiral galaxies. The
results of the generalised STY fitting for the late-type spirals are
superimposed for $z$=0.1, 0.3 and 0.5.}
\label{fig:elfsdm}
\end{figure}

\subsection{Summary and a Check}

\begin{figure}
\plottwo{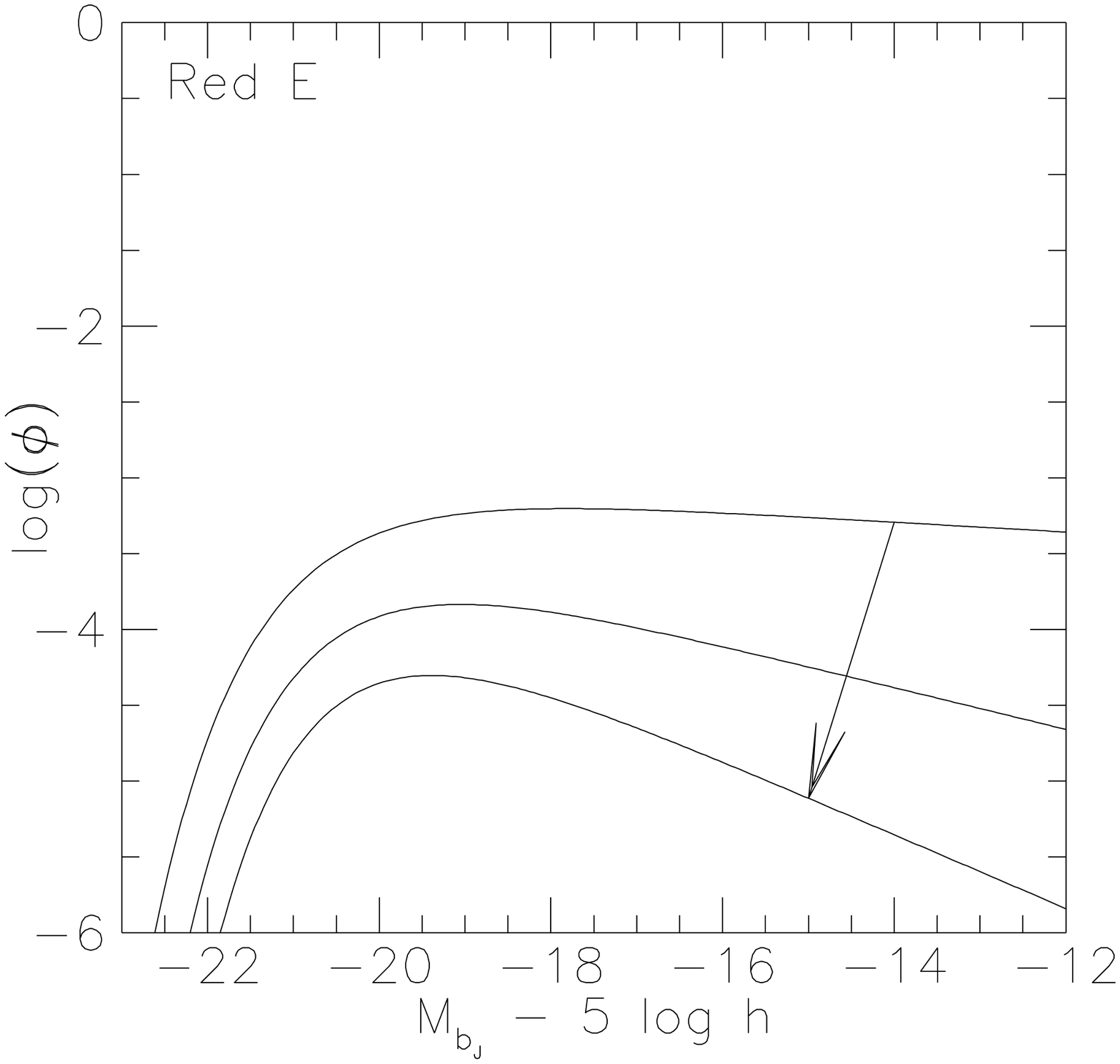}{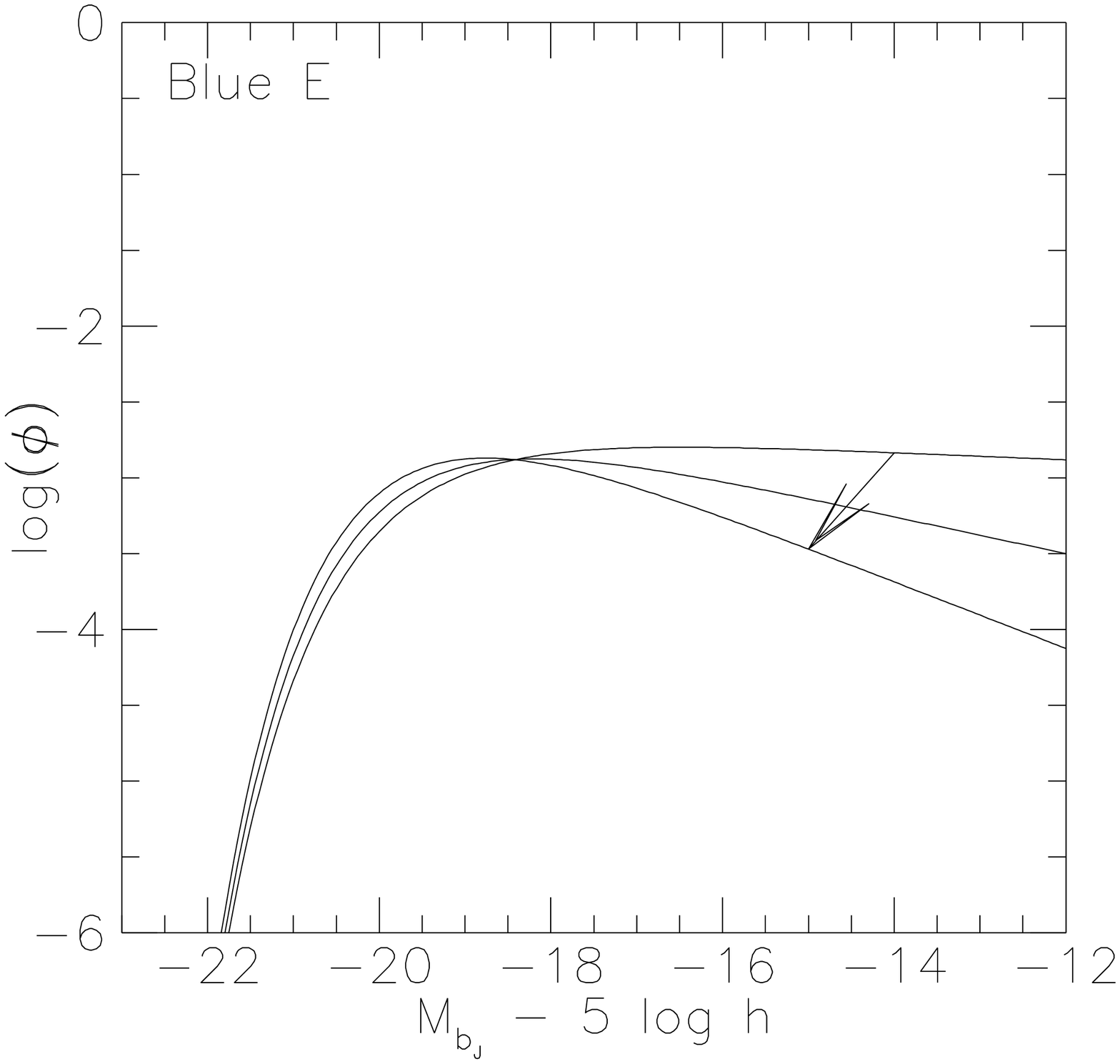}\\
\plottwo{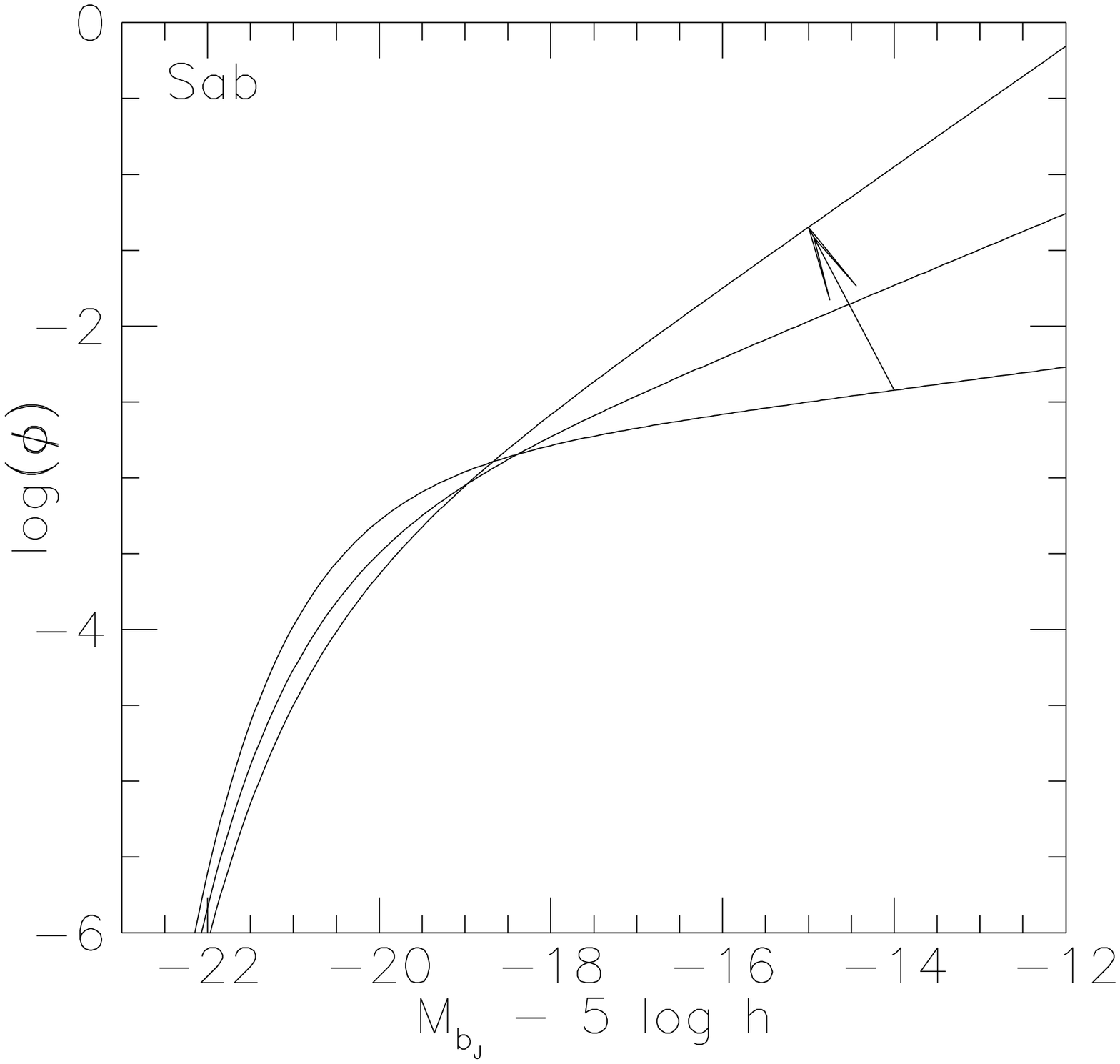}{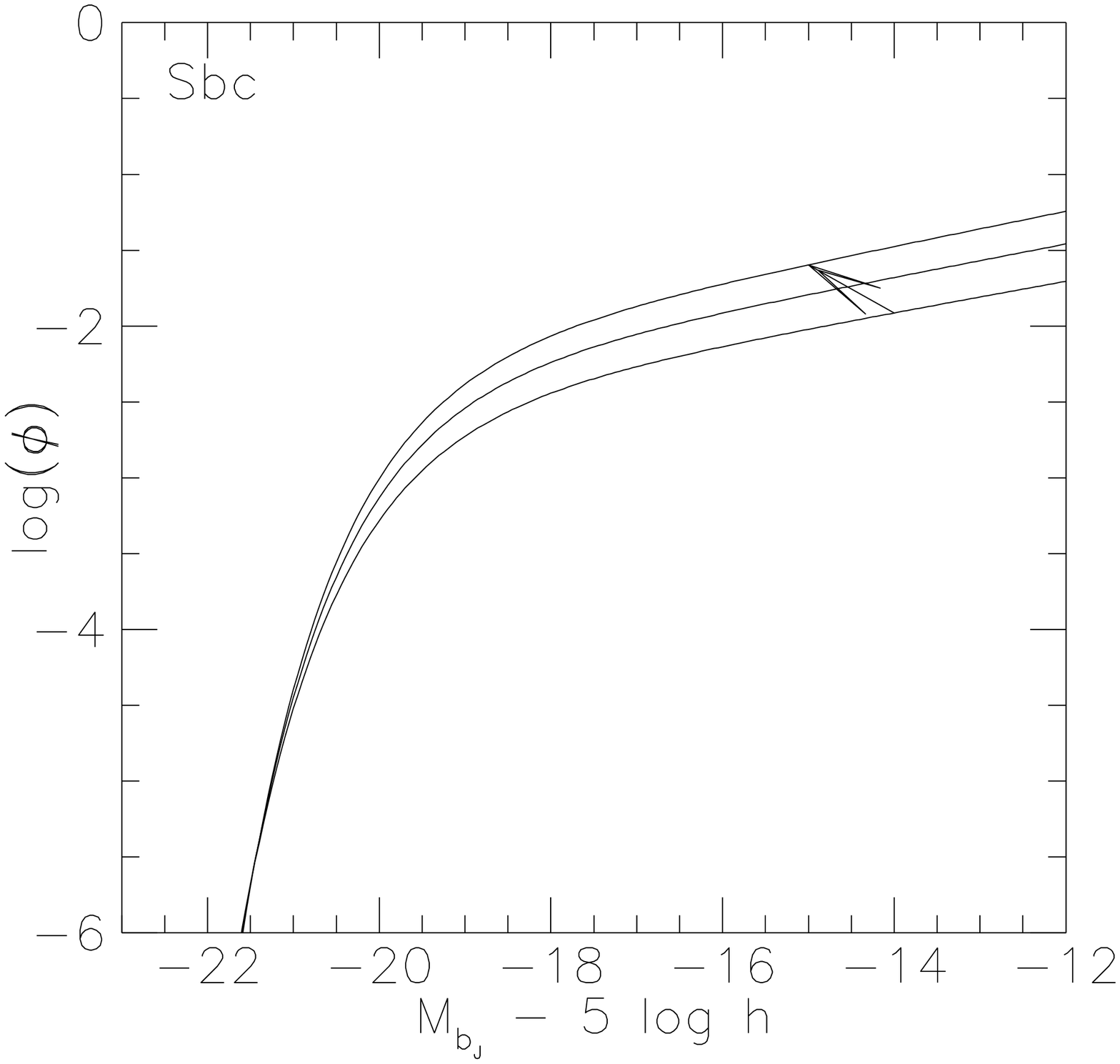}\\
\plottwo{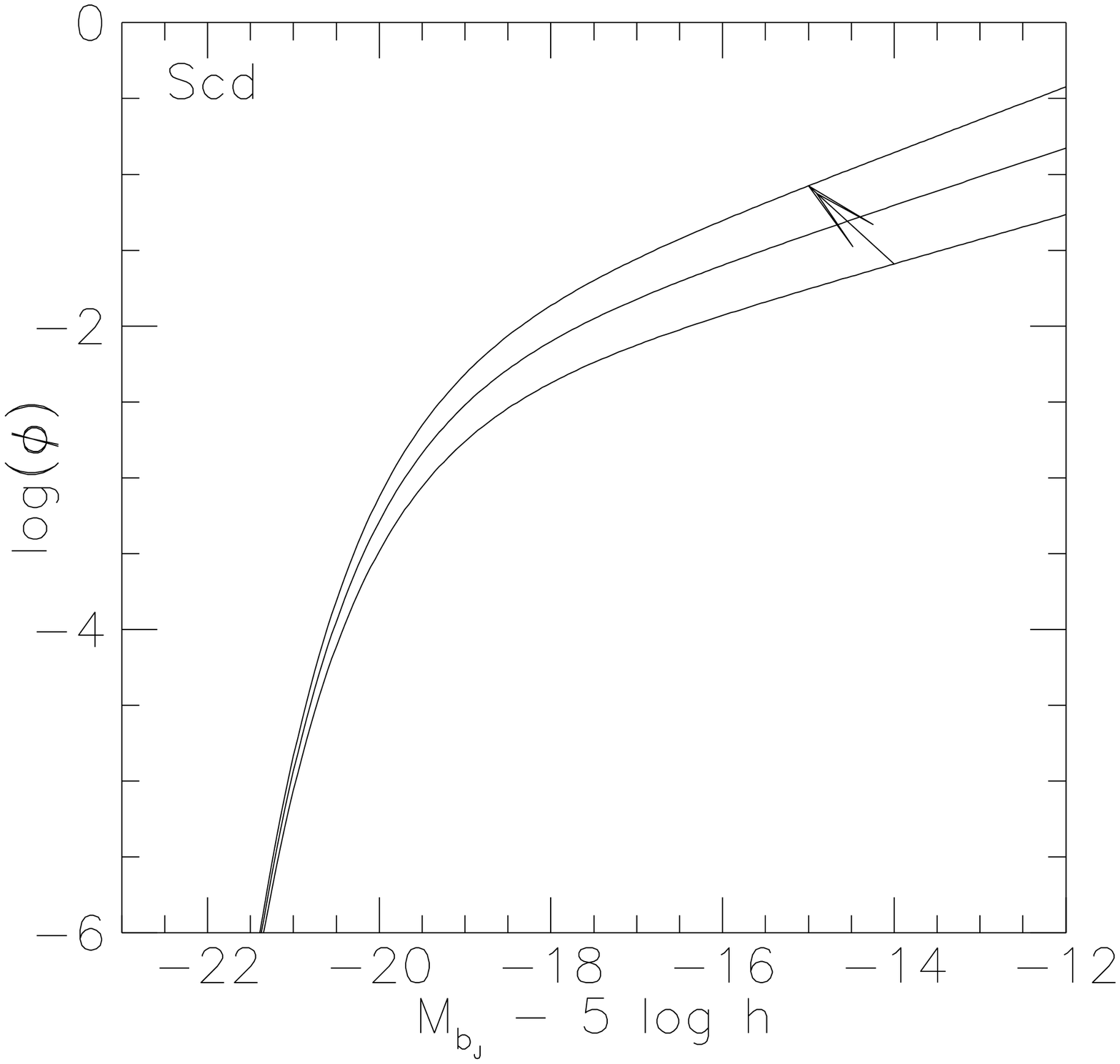}{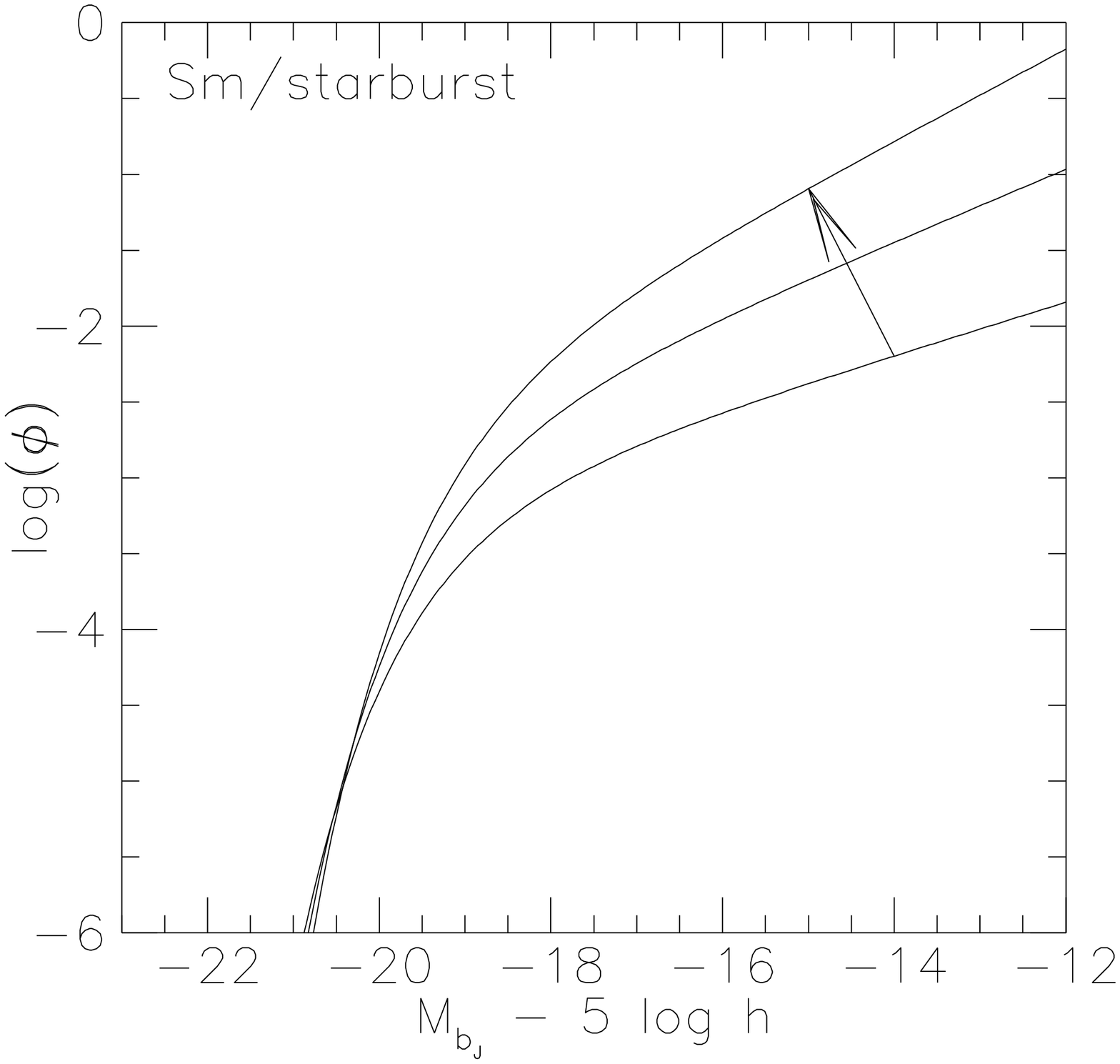}
\caption{The evolution of the luminosity function by spectral type. The
curves trace schematically the best-fitting, monotonically evolving
luminosity functions determined by the SSTY method at $z$=0.1, 0.3 and
0.5. Arrows indicate increasing redshift.}
\label{fig:allclasself}
\end{figure}

The evolution of the various galaxy types is summarised in pictorial
form in Figure~\ref{fig:allclasself}. This characterisation goes further 
than in the preceding sections and provides SSTY characterisations of 
the LF evolution for each of {\it six} spectral classes. With this level 
of subdivision the LFs are individually highly uncertain even within the
framework of the assumed evolutionary form (especially for the early-type 
class which is poorly represented at high redshift). Nonetheless the figure
illustrates that the earlier spectral types only weakly evolve whereas the 
rate of change is greatest for the later types. It is also clear that
the evolution takes the form of a steepening of the faint end of the LF 
with $L^*$ galaxies only being significantly affected at $z$$\sim$0.5. 
The parameters of the SSTY fits are provided in Table~\ref{tab:evpar}, 
although, in light of the above caveats, should clearly by used and 
interpreted with caution.

\begin{table*}
\caption{The evolution of the LF parameters.}
\label{tab:evpar}
\begin{tabular}{ccccccc}
Type & $\phi^*_0$ (Mpc$^{-3}$) & $\gamma_\phi$ & $M^*_0$ & 
$\gamma_L$ & $\alpha_0$ & $\gamma_\alpha$ \\
Red E & 1.62$\times$10$^{-3}$ & $-$6.15 & $-$20.7 & $-$1.77 & $-$1.05
& 1.31 \\
Blue E & 1.87$\times$10$^{-3}$ & 1.56 & $-$19.57 & $-$0.35 & $-$1.06
& 1.23 \\
Sab  & 2.19$\times$10$^{-3}$ & $-$4.44 & $-$20.0 & 0.92 & $-$0.99 & $-$2.01 \\ 
Sbc  & 2.80$\times$10$^{-3}$ & 2.89 & $-$19.4 & $-$0.37 & $-$1.25 & $-$0.07 \\ 
Scd  & 3.01$\times$10$^{-3}$ & 3.48 & $-$19.2 & $-$0.18 & $-$1.37 & $-$0.34 \\
Sdm/Starburst  & 0.50$\times$10$^{-3}$ & 6.61 & $-$19.0 & $-$0.97 & $-$1.36 & $-$0.79 
\end{tabular}
\end{table*}

\begin{figure}
\plotone{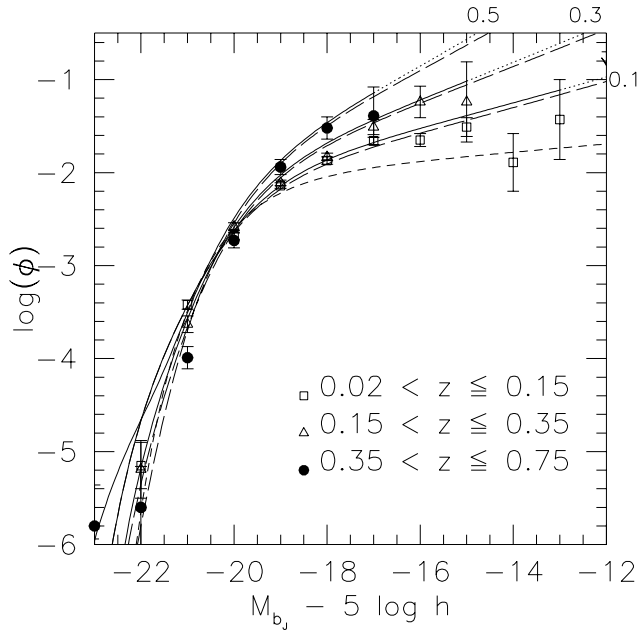}
\caption{Comparison of the SSTY and SSWML luminosity functions. The
sum of the SSTY LFs for the six classifications and the unclassified
galaxies (the solid curves) is compared with the LF for all the
galaxies in the survey as determined by the SSWML method. The Loveday
et~al.\ (1992) LF is shown as the short-dashed curve. Just below the
solid curves are long-dashed curves, which show the evolution of the
luminosity function with unclassified galaxies excluded from the
analysis. The SSTY results are given for $z$=0.1, 0.3 and 0.5.}
\label{fig:totalelf}
\end{figure}

As a final check, the SSTY LF fits for each spectral type are combined
to give the evolution of the overall luminosity function. This is
compared to the SSWML LFs for the whole sample in Figure~\ref{fig:totalelf} 
and the consistency check demonstrates that for the large combined
sample the two approaches yield consistent results. We conclude that
the evolutionary framework assumed in the SSTY fits is sufficiently
general to adequately match the observations.

\section{Discussion}
\label{sec:discuss}

The evolution of the LF by spectral type described in the previous
section considerably extends the analyses of previous studies, which
have either used [OII] EW (Paper~I) or red/blue colour (\cite{Lill95})
to make a simple distinction between the star-forming and quiescent
galaxy populations. Comparison of those studies with the more detailed
results presented here, however, shows broad agreement. The analyses
based here will pave the way for more comprehensive analyses that
will become possible in more extensive homogeneous faint spectral 
surveys.

In comparing the LF evolution of the red and blue galaxies in their
I$<$22 redshift survey, \jcite{Lill95} found that there was very
little change in the luminosity or number density of the red galaxies
over the whole redshift range 0$<$$z$$<$1, a point which we can
confirm both in terms of the derived LF and the mean spectral characteristics.
Amongst the blue galaxies \jcite{Lill95} found substantial evolution 
at $z$$>$0.5, so that by 0.5$<$$z$$<$0.75 the blue LF had brightened by 
about 1~mag. They were noncommittal about the evolution of the blue 
LF at lower redshifts, due to the small number of objects at $z$$<$0.2 
(35) with which to compare their 0.2$<$$z$$<$0.5 LF. Nonetheless, 
their Schechter function parameters indicated that the blue galaxies 
with $z$$<$0.5 (which are mostly at the high end of our redshift range) 
have a LF with a faint-end slope $\alpha$ between $-$1.3 and $-$1.6.
This is as steep or steeper than the value we obtain for low-redshift mid- to 
late-type spirals in our SSTY parametric analysis (see Table~3), but 
quite consistent with the values we find for these types at $z$$\ls$0.5. 
With a much larger sample at lower redshifts, our results not only 
show that significant evolution is occurring at $z$$<$0.5, but also 
demonstrate the marked type-dependence, with late-type spirals evolving 
more strongly than mid-type spirals. 

Many authors have previously postulated that late-type galaxies are the
prime movers in the evolution of the luminosity function as appears to 
be manifest in the number-magnitude relation. Their blue colours imply 
that they are least affected by $k$-corrections and thus can be observed 
to the largest distances for a given absolute magnitude. \jcite{BES} found 
that a luminosity function with a constant cutoff luminosity and a 
faint-end slope which increases with redshift is consistent with both 
the slope of the observed blue number counts and the redshift 
distribution. \jcite{Lace91} and \jcite{Trey94} both supported this finding, 
and conjectured that these changes are driven by blue, late-type galaxies 
which have disappeared since a redshift of $z$$\sim$0.2. \jcite{Lace91} 
propose several options for their disappearance. 

The excess galaxies could have produced more high-mass stars
proportionally than modern galaxies.  This bias toward high-mass stars
would make them brighter while stars were forming, but the galaxies
would quickly fade after the high-mass stars became supernova, leaving
few stars to be observed today. An alternative explanation is that
these galaxies have merged into the galaxies that we observe today.
\jcite{Whit90} and \jcite{Efst90m} argue that such an evolutionary
history is a natural consequence of the turnaround of larger and larger
mass scale with time in a hierarchical universe.  However, large
amounts of merging would leave traces in today's galaxies which are not
observed (\cite{Toth92}).  Furthermore, these excess galaxies are
unlikely to be the progenitors of today's galaxies as they are more
weakly clustered and more dense than galaxies today (\cite{Babu91}).
However, there is no convincing evidence in either our early-type
spectra or LFs for a significant build up of more massive early types
(c.f. \jcite{Dalc93}) as would be the case in the merger hypotheses.

\jcite{Babu91} propose a further explanation (elucidated also in
\cite{Efst92}). Although small mass haloes ($\sim 10^9 M_\odot$) collapse 
and virialise before the nascent haloes of $L^*$ galaxies, the UV flux
produced by quasars may keep the gas in small haloes ionised until
$z$$\sim$1. Only then can these small galaxies begin to form stars.
Such stars would form quickly as in a starburst galaxy. During 
the starburst these galaxies would appear irregular and their spectra
would be similar to those of present-day late-type spirals (i.e. dominated 
by short-term stellar activity). Supernovae would blow out the gas, halting
the star formation rate, and the galaxy would begin to fade (more quickly 
in the $B$-band than in the $K$-band).  \jcite{Babu91} also proposed that 
in low-pressure environments the gas may escape the galaxy entirely, while 
in intermediate and high-pressure regions, some gas might return
to the galaxy, possibly fuelling further events. The few galaxies
that could still be observed today would be in the high-pressure
regions clustered near more luminous galaxies.  Gigayears of phase 
mixing could transform these irregular starburst galaxies into today's
population of dwarf elliptical galaxies.  However, the vast majority
of these remnant irregulars would fade below detection limits. 
Although the clear proof of this picture would be the identification of 
an abundant population of local remnants, presumably via deep infrared
imaging, the short-term nature of star formation in the coadded spectra 
of the dominant evolutionary component provides some support for this picture.
With the larger databases anticipated in the forthcoming deep surveys,
it will hopefully become possible to tie the timescales observed
in the redshift evolution to those indicated by the detailed spectral
features in the coadded spectra.
 
\section{Conclusions}
\label{sec:conclude}

This paper investigates the evolution of the galaxy luminosity
function for galaxies of different spectral types. The analysis is
based on the Autofib redshift survey described in \jcite{Elli96} and
uses new techniques developed for spectral classification and for
recovering the luminosity function in the presence of both evolution
and clustering.

We classify the redshift survey galaxies into six spectral types using
a cross-correlation method that simulations show correctly assigns the
type for 80\% of the sample and is out by at most one type more than 90\%
of the time. These types compare well with those derived from the
galaxies' colours and, where available, with Hubble Space Telescope
morphologies.

We describe an extension of the step-wise maximum likelihood method
for estimating luminosity functions, rigorously testing the method
and comparing its performance to the standard \vm\ method. 
Significant advantages to the new SSWML method are found in reducing 
the sensitivity of the results to clustering. We also describe an 
extension to the \jcite{STY} method that enables us to recover the best 
fit to a parametric model for the luminosity function including its
evolutionary behaviour.

Applying these methods to the Autofib redshift survey we obtain the
following main results:

1.~There is no significant evolution in the mean spectrum or luminosity
function of early-type galaxies out to at least $z$$\sim$0.5. This
significantly constrains the continued production of such galaxies
(e.g. via mergers of star-forming galaxies over the last few Gyr).

2.~Early-type spiral galaxies show relatively modest evolution which
is well-characterised by a simple steepening of the faint end of their
luminosity function with redshift. Early spirals show little evidence
for evolution of either $L^*$ or $\phi^*$ over this redshift range.
Their spectra at all redshifts sampled are consistent with smooth
changes in the overall star formation history.

3.~Out to $z$$\sim$0.5, the overall evolution of the galaxy population
is dominated by the evolution of late-type spirals. Their luminosity
function not only steepens and brightens (evolution of both $\alpha$
and $L^*$) but there are also signs of significant density evolution
(a rapid increase in $\phi^*$). There appears to be a qualitative
change in the spectra of late type spirals with redshift. In addition
to a rise in the median [OII] equivalent width, a greater proportion of 
high redshift sources show Balmer line ratios indicative of short-term 
star formation. Coupled to the increase in star-formation rate at 
lower luminosities at fixed redshift, this implies that a given 
star-formation rate is found at higher redshift in higher 
luminosity objects.

\section*{ACKNOWLEDGEMENTS}

We acknowledge useful discussions with Len Cowie, Simon Lilly, Olivier
LeFevre, Donald Lynden-Bell and David Koo.  We thank Jarle Brinchmann
and Karl Glazebrook for assistance with analyses of the HST images.
JSH thanks the Marshall Aid Commission.  MMC acknowledges the
assistance of the Australian Academy of Science/Royal Society exchange
program. RSE and KGB acknowledge financial support from PPARC.


\begin{thebibliography}{}

\bibitem[\protect{Abraham et al.~\poy1996\pcy}]{Abra96}
Abraham R.G., Tanvir N.R., Santiago B.X., Ellis R.S., Glazebrook K.,
van den Bergh S., MNRAS, 279, L47.

\bibitem[\protect{Avni \& Bahcall~\poy1980\pcy}]{Avni80}
Avni Y., Bahcall J.N., 1980, ApJ, 235, 694

\bibitem[\protect{Babul \& Rees~\poy1991\pcy}]{Babu91}
Babul A., Rees M.J., 1991, MNRAS, 255, 346

\bibitem[\protect{Bender et~al.~\poy1993\pcy}]{Bend93}
Bender R., Burstein D., Faber S.M., 1993, ApJ, 411, 153

\bibitem[\protect{Binggeli, Sandage \& Tammann~\poy1988\pcy}]{Bing88}
Binggeli B., Sandage A., Tammann G.A., 1988, ARA\&A, 26, 509

\bibitem[\protect{Broadhurst, Ellis \& Shanks~\poy1988\pcy}]{BES}
Broadhurst T.J., Ellis R.S., Shanks T., 1988, MNRAS, 235, 827

\bibitem[\protect{Cho{\l}oniewski~\poy1986\pcy}]{Chol86}
Cho{\l}oniewski J., 1986, MNRAS, 223, 1

\bibitem[\protect{Colless et~al.~\poy1990\pcy}]{LDSS1}
Colless M.M., Ellis R.S., Taylor K., Hook R.N., 1990, MNRAS, 244, 408

\bibitem[\protect{Colless et~al.~\poy1993\pcy}]{Coll93} 
Colless M.M., Ellis R.S., Broadhurst T.J., Taylor K., Peterson B.A.,
1993, MNRAS, 261, 19

\bibitem[\protect{Cowie et~al.~\poy1991\pcy}]{Cowi91}
Cowie L.L., Songaila A., Hu E.M, 1991, Nature, 354, 460 

\bibitem[\protect{Dalcanton\poy1993\pcy}]{Dalc93}
Dalcanton J.J., 1993, ApJ, 415, L87

\bibitem[\protect{deVaucouleurs~\poy1977\pcy}]{deVa77}
deVaucouleurs, G., 1977, in Tinsley B.M. \& Larson R.B., ed.,
The Evolution of Galaxies and Stellar Populations, Yale University
Observatory, New Haven, p.43

\bibitem[\protect{Eales~\poy1993\pcy}]{Eale93}
Eales S., 1993, ApJ, 404, 51

\bibitem[\protect{Efstathiou~\poy1990\pcy}]{Efst90m}
Efstathiou, G., 1990, in Wielen R., ed., Dynamics and Interactions of
Galaxies, Springer, Berlin, p.2

\bibitem[\protect{Efstathiou~\poy1992\pcy}]{Efst92}
Efstathiou G., 1992, MNRAS, 256, 43P

\bibitem[\protect{Efstathiou, Ellis \& Peterson~\poy1988\pcy}]{EEP}
Efstathiou G., Ellis R.S., Peterson B.A., 1988, MNRAS, 232, 431

\bibitem[\protect{Ellis et al.~\poy1996\pcy}]{Elli96}
Ellis R.S., Colless M.M., Broadhurst T.J., Heyl J.S., Glazebrook K., 
1996, MNRAS, 280, 235, (Paper~I)

\bibitem[\protect{Felten~\poy1976\pcy}]{Felt76}
Felten J.E., 1976, ApJ, 207, 700

\bibitem[\protect{Glazebrook et al.~\poy1995a\pcy}]{Glaz95a}
Glazebrook K., Ellis R.S., Colless M.M., Broadhurst T.J.,
Allington-Smith J.R., Tanvir N.R., 1995a, MNRAS, 273, 157

\bibitem[\protect{Glazebrook et al.~\poy1995b\pcy}]{Glaz95b}
Glazebrook K., Ellis R. S., Santiago B. X., Griffiths R. E. 1995b,
MNRAS, 275, L19

\bibitem[\protect{Heyl~\poy1994\pcy}]{Heyl94}
Heyl J.S., 1994, M.Sc.\ Thesis, University of Cambridge

\bibitem[\protect{Kennicutt~\poy1992a\pcy}]{Kenn92a}
Kennicutt R., 1992a, ApJS, 79, 255

\bibitem[\protect{Kennicutt~\poy1992b\pcy}]{Kenn92b}
Kennicutt R., 1992b, ApJ, 388, 410

\bibitem[\protect{King \& Ellis~\poy1985\pcy}]{King85}
King C.R., Ellis R.S., 1985, ApJ, 288, 456

\bibitem[\protect{Kinney et~al.~\poy1993\pcy}]{Kinn93}
Kinney, A. L., Bohlin, R. C., Calzetti, D., Panagia, N. \& 
Wyse, Rosemary F. G 1993, ApJS, 86, 5

\bibitem[\protect{Koo \& Kron~\poy1992\pcy}]{Koo92}
Koo D.C., Kron R.G., 1992, ARA\&A, 30, 613

\bibitem[\protect{Lacey \& Silk~\poy1991\pcy}]{Lace91}
Lacey C.G., Silk J., 1991, ApJ, 381, 14

\bibitem[\protect{Lilly et~al.~\poy1995\pcy}]{Lill95}
Lilly S.J., Tresse L., Hammer F., Crampton D., Le Fevre O., 1995, ApJ,
455, 108

\bibitem[\protect{Lin et~al.~\poy1996\pcy}]{Lin96}
Lin H., Kirshner R.P., Shectman S.A., Landy S.D., Oemler A., Tucker
D.L., Schechter P.L., 1996, ApJ, in press

\bibitem[\protect{Lonsdale \& Chokshi~\poy1993\pcy}]{Lons93}
Lonsdale C.J., Chokshi A., 1993, AJ, 105, 1333

\bibitem[\protect{Loveday et~al.~\poy1992\pcy}]{Love92}
Loveday J., Peterson B.A., Efstathiou G., Maddox S.J., 1992, ApJ, 390,
338

\bibitem[\protect{Lynden-Bell~\poy1971\pcy}]{Lynd71}
Lynden-Bell D., 1971, MNRAS, 155, 95

\bibitem[\protect{Marshall et~al.~\poy1983\pcy}]{Mars83}
Marshall H.L., Avni Y., Tananbaum N., Zamorani G., 1983, ApJ, 269, 35

\bibitem[\protect{Marzke et~al.~\poy1994a\pcy}]{Marz94}
Marzke R.O., Geller M.J., Huchra J.P., Corwin H.G., 1994, AJ, 108, 437

\bibitem[\protect{Pence~\poy1976\pcy}]{Penc76}
Pence W., 1976, ApJ, 203, 39

\bibitem[\protect{Peterson et al.~\poy1985\pcy}]{Pete85}
Peterson B.A., Ellis R.S., Bean A.J., Efstathiou G.P., Shanks T.,
Fong R., Zou Z-L., 1985, MNRAS, 221, 233.

\bibitem[\protect{Sandage, Tammann \& Yahil~\poy1979\pcy}]{STY}
Sandage A., Tammann G.A., Yahil A., 1979, ApJ, 232, 352

\bibitem[\protect{Saunders et al.~\poy1990\pcy}]{Saun90}
Saunders W., Rowan-Robinson M., Lawrence A., Efstathiou G., Kaiser N.,
1990, MNRAS, 242, 318

\bibitem[\protect{Schmidt~\poy1968\pcy}]{Schm68}
Schmidt M., 1968, ApJ, 151, 393

\bibitem[\protect{Taylor~\poy1995\pcy}]{Tayl95}
Taylor K., 1995, BAAS, 186, \#44.06

\bibitem[\protect{Toth \& Ostriker~\poy1992\pcy}]{Toth92}
Toth G., Ostriker J.P., 1992, ApJ, 389, 5

\bibitem[\protect{Treyer \& Silk~\poy1994\pcy}]{Trey94} 
Treyer M.A., Silk J., 1994, ApJ, 436, L143

\bibitem[\protect{White~\poy1990\pcy}]{Whit90}
White S.D.M., 1990, in Wielen R., ed., Dynamics and Interactions of
Galaxies, Springer, Berlin, p.380

\bibitem[\protect{Zabludoff et al.~\poy1996\pcy}]{Zabl96}
Zabludoff A.I. et al. 1996, ``The Environment of E+A Galaxies,''
LANL Preprint: astro-ph/9512058, ApJ,
submitted.

\bibitem[\protect{Zaritsky, Zabludoff \&
Willick~\poy1995\pcy}]{Zari95}
Zaritsky, D., Zabludoff, A.I., Willick, J.A. 1995, AJ, 110, 1602.

\end{thebibliography}
\end{document}